\documentclass[a4paper,11pt]{article}
\usepackage{jheppub}
\usepackage{CJK}
\usepackage{tikz}
\usepackage{float,graphbox,makecell,tablefootnote}
\usepackage[caption=false]{subfig}
\usepackage{xcolor}
\usepackage{longtable}
\usepackage{multirow}
\usepackage{booktabs}
\allowdisplaybreaks[4]
\usepackage{diagbox}
\usepackage{alltt}
\usepackage[theorems]{tcolorbox}

\usepackage{tabularray}
\usepackage{xparse}

\usetikzlibrary{decorations.pathreplacing, calc, arrows.meta}

\tikzset{point/.style={circle,inner sep=2pt}, axisStyle/.style={line width=2pt, line cap=round}}

\definecolor{lightgreen}{rgb}{0.56, 0.69, 0.19}
\definecolor{lightblue}{rgb}{0.36, 0.51, 0.71}
\definecolor{lightyellow}{rgb}{0.88, 0.61, 0.14}
\definecolor{darkgreen}{rgb}{0.6, 0.6, 0.35}
\definecolor{lightred}{rgb}{0.99, 0.36, 0.02}
\definecolor{box1}{rgb}{0.46, 0.6, 0.45}
\definecolor{box2}{rgb}{0.62, 0.56, 0.43}
\definecolor{box3}{rgb}{0.72, 0.65, 0.17}

\definecolor{darkred}{rgb}{0.5,0.0,0.0}
\definecolor{darkblue}{rgb}{0.0,0.0,0.9}
\definecolor{darkerblue}{rgb}{0.0,0.0,0.5}
\definecolor{darkgreen}{rgb}{0.0,0.5,0.0}
\definecolor{black}{rgb}{0.0,0.0,0.0}
\definecolor{brown}{rgb}{0.6,0.4,0.2}
\newcommand{\red}{\color{darkred}}
\newcommand{\blue}{\color{darkerblue}}

\newcommand{\black}{\color{black}}
\newcommand{\vev}[1]{\langle #1 \rangle}
\newcommand{\sqb}[1]{[ #1 ]}
\newcommand{\ab}[1]{\langle #1 \rangle}

\usepackage[hang]{footmisc}
\makeatletter
\ifFN@para
\else
  \long\def\@makefntext#1{%
    \ifFN@hangfoot
      \bgroup
      \setbox\@tempboxa\hbox{%
        \ifdim\footnotemargin>0pt
          \hb@xt@\footnotemargin{\@makefnmark\hss}%
        \else
          \@makefnmark\hskip-\footnotemargin      
        \fi
      }%
      \leftmargin\wd\@tempboxa
      \rightmargin\z@
      \linewidth \columnwidth
      \advance \linewidth -\leftmargin
      \parshape \@ne \leftmargin \linewidth
      \footnotesize
      \@setpar{{\@@par}}%
      \leavevmode
      \llap{\box\@tempboxa}%
      \parskip\hangfootparskip\relax
      \parindent\hangfootparindent\relax
    \else
      \parindent1em
      \noindent
      \ifdim\footnotemargin>\z@
        \hb@xt@ \footnotemargin{\hss\@makefnmark}%
      \else
        \ifdim\footnotemargin=\z@
          \llap{\@makefnmark}%
        \else
          \llap{\hb@xt@ -\footnotemargin{\@makefnmark\hss}}%
        \fi
      \fi
    \fi
    \footnotelayout#1%
    \ifFN@hangfoot
      \par\egroup
    \fi
  }
\fi
\makeatother

\usepackage{slashed}

\usepackage{tikz}

\def\cE{\mathcal{E}}

\def\cL{\mathcal{L}}

\def\cN{\mathcal{N}}
\def\cO{\mathcal{O}}

\def\nn{\nonumber}

\def\r{\rangle}

\newcommand{\sblue}[1]{\color{lightblue}{#1}}
\def\FF{\mathbb{F\bar{F}}}
\def\FFint{\mathrm{F\bar{F}}}

\makeatletter
\newcommand{\subalign}[1]{
    \vcenter{
    \Let@ \restore@math@cr \default@tag
    \baselineskip\fontdimen10 \scriptfont\tw@
    \advance\baselineskip\fontdimen12 \scriptfont\tw@
    \lineskip\thr@@\fontdimen8 \scriptfont\thr@@
    \lineskiplimit\lineskip
    \ialign{\hfil$\m@th\scriptstyle##$&$\m@th\scriptstyle{}##$\hfil\crcr
    #1\crcr}}
}

\newcommand{\raisemath}[1]{\mathpalette{\raisem@th{#1}}}
\newcommand{\raisem@th}[3]{\raisebox{#1}{$#2#3$}}
\makeatother

\input{tikz_figs.tex}

\begin{document}

\begin{CJK*}{UTF8}{}
\CJKfamily{gbsn}

\title{Bootstrapping form factor squared in ${\cal N}=4$ super-Yang-Mills}

\author[a,b,c]{Song He}\emailAdd{songhe@itp.ac.cn}
\author[a,d]{Xiang Li}\emailAdd{lixiang@itp.ac.cn}
\author[e,f]{Jingwen Lin}\emailAdd{linjingwen@sjtu.edu.cn}
\author[a,d]{Jiahao Liu}\emailAdd{liujiahao@itp.ac.cn}
\author[e,f]{Kai Yan}\emailAdd{yan.kai@sjtu.edu.cn}

\affiliation[a]{Institute of Theoretical Physics, Chinese Academy of Sciences, Beijing 100190, China}
\affiliation[b]{School of Fundamental Physics and Mathematical Sciences, Hangzhou Institute for Advanced Study and ICTP-AP, UCAS, Hangzhou 310024, China}
\affiliation[c]{Peng Huanwu Center for Fundamental Theory, Hefei 230026, China}
\affiliation[d]{School of Physical Sciences, University of Chinese Academy of Sciences, No.19A Yuquan Road, Beijing 100049, China}
\affiliation[e]{State Key Laboratory of Dark Matter Physics, Shanghai Key Laboratory for Particle Physics and Cosmology, Key Laboratory for Particle Astrophysics and Cosmology (MOE), School of Physics and Astronomy, Shanghai Jiao Tong University, Shanghai 200240, China}
\affiliation[f]{Institute of Nuclear and Particle Physics (INPAC), Shanghai Jiao Tong University, Shanghai 200240, China}

\abstract{We propose a bootstrap program for the {\it form factor squared} with operator ${\rm tr}(\phi^2)$ in maximally supersymmetric Yang-Mills theory in the planar limit, which plays a central role for perturbative calculations of important physical observables such as energy correlators. The tree-level $N$-point form factor (FF) squared can be obtained by cutting $N$ propagators of a collection of two-point ``master diagrams" at $(N{-}1)$ loops: for $N=3,4,5,6$ there are merely $1, 2, 4, 13$ topologies of such diagrams respectively, and their numerators are strongly constrained by power-counting (including ``no triangle" property) and other constraints such as the ``rung rule". Moreover, these two-point diagrams provide a ``unification" of FF squared at different numbers of loops and legs, which is similar to extracting (planar) amplitude squared from vacuum master diagrams (dual to $f$-graphs): by cutting $2\leq n<N$ propagators, one can also extract the planar integrand of $n$-point FF squared at $(N-n)$ loops, thus our results automatically include integrands of 2-point (Sudakov) FF up to four loops (where the squaring is trivial), 3-point FF squared up to three loops, and so on. Our ansatz is completely fixed using soft limits of (tree and loop) FF squared and the multi-collinear limit which reduces it to the splitting function, without any other inputs such as unitarity cuts. This method opens up the exciting possibility of a {\it graphical bootstrap} for FF squared for higher $N$ (which contains {\it e.g.} planar Sudakov FF to $N{-}2$ loops) similar to that for the amplitude squared via $f$-graphs. We also comment on applications to the computation of leading order energy correlators where new structures are expected after performing phase-space integrations.
}

\maketitle
\end{CJK*}

\section{Introduction}
Recent years have witnessed enormous progress in the study of scattering amplitudes and form factors {\it etc.} in perturbative Quantum Field Theory: a driving force has been discoveries of hidden simplicity and new structures within the ``theoretical data'' resulting from rather hard computations, which often lead to deep insights, powerful new computational tools, and even more such discoveries. Compared to scattering amplitudes, physical cross-section level observables relevant for colliders are much less studied but similar remarkable simplicity and new structures in them are waiting to be revealed. 
An interesting class of such physical observables are multi-point correlation functions of energy flux~\cite{Basham:1977iq, Basham:1978bw,Basham:1978zq, Basham:1979gh,Chen:2020vvp}, denoted as Energy Correlators (EC), which are infrared finite~\cite{Kinoshita:1962ur, Lee:1964is} and can be directly measured in experiment~\cite{OPAL:1990reb,ALEPH:1990vew,SLD:1994yoe, Komiske:2022enw, Chen:2022swd, CMS:2023wcp,CMS:2024mlf, Tamis:2023guc}, making them ideal candidates for explorations. The most symmetric of all four-dimensional gauge theories, the ${\cal N}=4$ supersymmetric Yang-Mills (SYM) theory, has played a central role for all these studies, with quite a lot of new results for perturbative computations of EC~\cite{Hofman:2008ar,Belitsky:2013ofa,Henn:2019gkr,Yan:2022cye,Chicherin:2024ifn,He:2024hbb}. Recently, the study of multi-point EC in multi-collinear limits~\cite{Chen:2019bpb,Chicherin:2024ifn,He:2024hbb} has motivated a more systematical computation of {\it amplitude squared} or the {\it splitting functions} : at leading order, the $n$-point EC in collinear limit is given by the phase-space integral of the universal $1\to n$ {\it splitting functions}~\cite{Amati:1978wx,Amati:1978by,Ellis:1978sf,Catani:1998nv}, which are given by $(n{+}3)$-point tree amplitude squared in a local form~\cite{He:2024hbb}. In the planar limit, the most efficient way for computing amplitude squared/splitting function turns out to be from light-like limits of the four-point correlator via $f$-graphs: given the very recent advance to twelve loops~\cite{Bourjaily:2025iad,He:2024cej}, we have $16$-point amplitude squared ($1\to 13$ splitting functions) at hand which in principle provides us access to $13$-point EC in the multi-collinear limit!

On the other hand, we have not accumulated much theoretical data for EC integrands beyond the multi-collinear limit, which can be extracted from the more general {\it form-factor (FF) squared}: in multi-collinear limit, the $(n{+}1)$-point FF squared with operator ${\rm tr} (\phi^2)$ reduces to $1\to n$ splitting functions, but it gives $n$-point EC for general kinematics after phase-space integrals. In fact, these FF squared are key ingredients to leveraging perturbative calculations of various physical observables, with applications including energy correlators, infrared subtraction and multi-soft emission factors {\it etc.}. From a more formal perspective, it is also highly desirable to study FF squared since they are simpler objects (compared to FF themselves): the information on helicity (or Grassmann variables) of all external states is gone after gluing, thus they only depend on Mandelstam variables; both the tree-level FF squared and their loop integrands are rational functions of Lorentz products of (external and loop) momenta with simple, physical poles only. For the planar amplitude squared, not only do these rational functions manifest the dual conformal invariance (DCI) of the theory~\cite{Drummond:2006rz,Drummond:2008vq,Brandhuber:2008pf}, but integrands with different multiplicities and loops are nicely unified in the integrand of four-point correlator~\cite{Eden:2012tu,Eden:2010zz,EDEN2012450,Ambrosio:2013pba}, which exhibits a hidden permutation symmetry manifested by the $f$-graphs~\cite{Eden:2011we}. In contrast, very little is known on the FF side: it is unclear if there exists a duality between (light-like limits of) certain correlator and FF squared at integrand level, or what symmetries the final result should have, or if their integrands at different loops can be unified via some analog of $f$-graphs. The main goal of our study is to obtain explicit, compact expressions for FF squared, which will then allow us to reveal hidden structures and properties similar to amplitude squared. 

Note that for both amplitudes and FF, one can of course square the results from {\it e.g.} BCFW recursion or similar methods~\cite{Brandhuber:2010ad,Brandhuber:2011tv,Penante:2014sza,Bork:2012tt,Bork:2014eqa}, but the results are plagued with spurious poles and it becomes a daunting task for simplifying them to a local, usable form. Instead we will propose to directly derive such a form by a {\it bootstrap} method, which can be viewed as a generalization of the bootstrap for amplitude-squared based on $f$-graphs. As we will review shortly in Sec.~\ref{sec:FFsquare}, the tree-level $N$-point form factor squared, which we denote as $\FF^{(0)}_N$, can be obtained by gluing together two tree-level FF along their $N$ legs: diagrammatically they receive contributions from two-point diagrams (also called propagator diagrams) at $(N{-}1)$ loops, where the two external legs denote the operators (both with momentum $q$) (see Fig.~\ref{fig:cutn}). We will call such diagrams where all propagators are put off shell as ``master diagrams", and they can be viewed as a special version of generalized unitarity cuts studied in various contexts (see~\cite{Carrasco:2015iwa,Bern:2019prr,Yang:2019vag}, and references therein). For simplicity, we will focus on the planar limit ($N_c \to \infty$) where we only need a given color ordering and contributions from planar master diagrams (except for the insertion of operators): we expect that this general method still applies to non-planar FF but our goal here is to take the first step in extending the success of {\it planar} amplitude squared to FF squared. 

\begin{figure}[htbp]
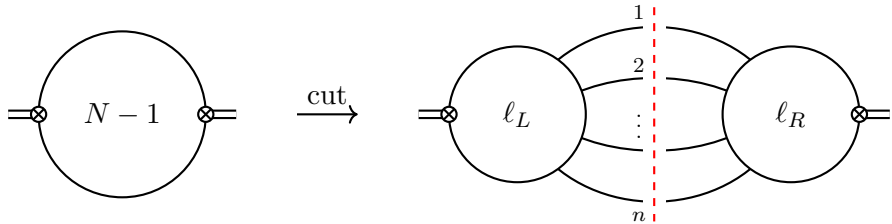

\centering
\ffcut
\caption{Two-point (or propagator) diagrams where the two external legs denote the operators (both with momentum $q$) at $(N{-}1)$ loops. Cutting $n$ propagators of the master diagram, which separates the two operators, gives the contribution where the two FFs have $\ell_L$ and $\ell_R$ loops. We have $ N = \ell_L + \ell_R + n $.}
\end{figure}
\label{fig:cutn}

\begin{figure}[htbp]
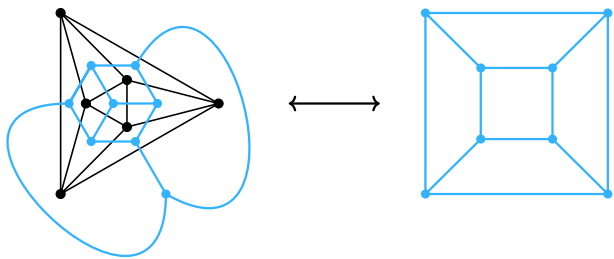

    \centering
    \vspace{-2em}
    \hspace{-6em}\scalebox{.6}{\fgraphdual}
    \vspace{-5em}
    \caption{The $6$-point $f$-graph is drawn by black lines, and its  Poincar{\'e} dual, \textit{i.e.} the vacuum diagram, is drawn in blue lines.}
    \label{fig:fgraphdual}
\end{figure}

Recall that for amplitude squared, the corresponding master diagrams are ``vacuum diagrams", which in the context of planar ${\cal N}=4$ SYM are the Poincar{\'e} dual of the $f$-graphs, and an example of $6$-point is shown in Fig.~\ref{fig:fgraphdual}. These $f$-graphs in the dual space have been proven to be the key for the bootstrap of amplitude squared: in addition to ``unifying" various amplitude squared, there is a unique rational function symmetric in all dual points, $x_1, x_2, \cdots, x_N$, known as ``the integrand" of the four-point correlator (or Born-level correlator with $N{-}4$ chiral Lagrangian insertions), which in various light-like limits reduce to the integrand of amplitude squared (with $n{+}\ell=N$). We emphasize that while one can also rewrite the two-point master diagrams for FF squared in dual space (on a cylinder with periodicity), currently we do not know of such a unique rational function unifying all FF squared (at integrand level), but rather we have an equivalent class of ``integrands" which indeed has ambiguities related to the definition of loop momentum (or the periodicity of the dual points). Of course this does not prevent us from studying such rational functions from the master diagrams, which by cutting internal propagators (or taking light-like limits of ``periodic'' cycles in dual space) nicely reduce to equivalent classes of ``integrands" of FF squared (with the exception that the tree-level FF squared has no such ambiguity). As a proof of concept, we will show how to bootstrap these functions from two-point master diagrams in momentum space, and leave a more systematic and efficient bootstrap program similar to that based on $f$-graphs to the future. 

As mentioned above, $\FF^{(0)}_N$ has no ambiguity and it can be simply obtained by summing over all ways of cutting $N$ propagators from the two-point diagrams, which automatically takes the local form. While a direct Feynman-diagram computation quickly becomes too complicated, our main point is that at least up to the $N=6$ case, it is straightforward to {\it bootstrap} these rational functions associated with the master diagrams, {\it i.e.} fixing their numerators and coefficients, using physical constraints. These constraints include the soft limit which connects $\FF_N^{(0)}$ to $\FF_{N{-}1}^{(0)}$ (it also connects the loop integrand $\FFint_N^{(\ell)}$ to $\FFint_{N{-}1}^{(\ell)}$), as well as the multi-collinear limit which reduces it to the splitting functions. For planar $\FF_N^{(0)}$, we find only $1, 2, 4, 13$ top topologies of master diagrams for $N=3,4,5,6$, and we will see in Sec.~\ref{sec:bootstrap} that the corresponding ansatz is rather restrictive based on the familiar power-counting of ${\cal N}=4$ SYM and {\it e.g.} the ``rung rule". Note that when we rewrite a two-point diagram in dual space, in principle there could be different dual graphs corresponding to different {\it embeddings} of the same two-point diagram, each of which contributes individually to the result. Hence when considering in the dual space, the counting of top topologies for $N=3,4,5,6$ increases to $1,3,6,21$. However, this will not actually increase the size of the ansatz, since all these different dual graphs corresponding to the same two-point diagram must share the same coefficient.

Remarkably, by cutting $n<N$ propagators of these two-point master diagrams which separate the two operators on two sides, we automatically obtain (equivalent classes of) loop integrands $\FFint_n^{(\ell)}$ at $\ell:=N-n$ loops. This can be argued in general based on Cutkosky rule, but it is very satisfying to see how precisely everything works out (including all symmetry factors) so that $\FFint_n^{(\ell)}$ for $n=2,\cdots, N$ at $\ell=N{-}2, \cdots, 0$ loops are all unified in these master diagrams just as the amplitude squared in $f$-graphs. More precisely, from how the diagrams are cut into two parts, we obtain contributions where the two FF have $\ell_L$ and $\ell_R$ loops (with $\ell_L + \ell_R=\ell$, schematically shown in Fig.~\ref{fig:cutn}): for $n=2$ if we single out the part with $\ell_L=\ell$ (thus $\ell_R=0$), we obtain the $\ell$-loop integrand for Sudakov form factors (as the gluing of a $2$-point tree is trivial). We emphasize again that these integrands are not like planar integrands of amplitude (squared) which are unique rational functions of dual points, but they should be viewed as equivalent classes up to relabeling of loop momenta/periodicity of the dual points. We will present these new results for both trees and loop integrands of FF, including an ancillary file that records all explicit results in the dual space for tree-level FF squared and loop integrands for $N\leq 6$.

Finally, we discuss applications of FF to energy correlators in Sec.~\ref{sec:EEC}. 
In the context of EC, periodic dual space turns into a cycle parametrized in terms of energies and angles.    
The $N$-point EC are iterated integrals in terms of these on-shell parameters, whose  $\ell$-loop integrands can be translated directly from $\FFint_{N+1+\ell-m}^{(m)},\, \, m\leq \ell$ (note that this multiplicity $N$ should not be confused with the $N$ above).  
Unlike automated procedures for evaluating Feynman loop integrals, techniques for the on-shell phase-space integrations are currently underdeveloped. 
We carry on a proof-of-principle study on the function space of $N$-point EC at the leading order (LO), by inspecting the incidence relations among the singularity surfaces that emerge in $\FFint_{N{+}1}$.   
The formalism for FF organized in terms of master diagrams streamlines the classification of integral topologies. 
In the case of polylogarithmic sectors, we always encounter a single quadratic hypersurface intersecting with hyperplanes.  This enables us to construct finite {\it dlog forms} \cite{Arkani-Hamed:2010pyv} for the EC by means of twisted intersection theory \cite{Mastrolia:2018uzb,Mizera:2019vvs,Caron-Huot:2021xqj,Caron-Huot:2021iev,De:2023xue}.  In general, one may have up to two quadratic hypersurfaces intersecting each other.
A preliminary discussion is given to analyze the structure of beyond-polylogarithmic integrals.   
These results open up prospects toward higher loop integrated results via either integral reduction in parameter space~\cite{Chetyrkin:1981qh,Lee:2014tja,Chen:2019mqc,Artico:2023jrc} and differential equations \cite{Kotikov:1990kg,Kotikov:1991pm,Remiddi:1997ny,Henn:2013pwa,Muller-Stach:2012tgj,Adams:2017tga,Dlapa:2020cwj,Dlapa:2022wdu} or a direct bootstrap method.

\subsection{Form factor squared: definition and properties}\label{sec:FFsquare} 

We consider the tree-level $n$-point form factor for the scalar half-BPS operator $\cO={\rm tr}(\phi^2)$ in $\cN=4$ super Yang-Mills theory, which corresponds to the lowest component of chiral stress-tensor supermultiplet \cite{Chicherin:2014uca}. 
\begin{align}
(2\pi)^4 \delta (q- \sum_{i=1}^n p_i) \, F_{n} (\cO )   \equiv  \int d^4 x\,  e^{i q \cdot x}\langle \Phi(1) \cdots \Phi(n) | \cO (x)| 0 \rangle
\end{align} 

Let $F^{(\ell)}_{n,k}$  be the color ordered $n$-point $\ell$-loop $\rm N^{\it k}MHV$  form factor. Let us define $\FF^{(\ell)}_{n,k}$ to be the $\ell$-loop color ordered product of form factor summed over on-shell super states, 
\begin{align}
\FF^{(\ell)}_{n,k}  & =  \sum_{\ell_L+\ell_R=\ell}\mathcal{C}_n  \int  \prod_{i=1}^n d^4 \eta_i  d^4 \bar{\eta}_i  \, e^{-\sum_i \bar \eta_i  \eta_i  }   F^{(\ell_L)}_{n,k} (\eta_i, \lambda_i )  \bar{F}^{(\ell_R)}_{n,k} (\bar{\eta}_i, \tilde{\lambda}_i), \quad   \\
& {\text{where}} \quad  \mathcal{C}_n = {\rm tr}(T^{a_1}  \cdots T^{a_n})    {\rm tr} (T^{a_n}  \cdots T^{a_1}).  \notag  
\end{align}
Taking the sum over helicity configurations, 
\begin{align}
\FF^{(\ell)}_{n} = \sum_{k=0}^{n-2} \FF^{(\ell)}_{n,k}  ,  \quad  \text{where} \quad  \FF^{(\ell)}_{n,k} =  \FF^{(\ell)}_{n,n-k-2}   .
\end{align} 
For $n \leq 5$, $\FF_n$ is the module square of form factor in full color. Starting at $n=6$, interference between $F_{n,k}$ with different color orderings does not vanish. They would contribute to the quartic Casimir color structure, which is subleading in $N_c$ \cite{Boels:2012ew}. We consider only the planar $N_c\rightarrow\infty$ limit, thus subleading contributions can be omitted. The color-ordered product $\FF_n$ contains the full leading color contributions.

As shown previously in Fig.~\ref{fig:cutn}, the planar $(n=N)$-point tree-level form factor squared $\FF^{(0)}_N$ can be obtained by cutting $N$ propagators of the summation of a collection of two-point $(N-1)$-loop master diagrams $\mathfrak{M}_N$, which corresponds to an equivalent class of integrands will be denoted by $\mathcal{I}_N$. This $\mathcal{I}_N$ automatically yields lower-point {\it loop integrand} for $\FF_n^{(\ell)}$, denoted by $\FFint_n^{(\ell)}$\footnote{Note that at $\ell=0$, we have $\FFint_n^{(0)} \equiv \FF_n^{(0)}$. Since the integrand $\FFint_n^{(\ell)}$ is the main topic of this paper, we will stick to the notation $\FFint_{N}^{(0)}$ from now on.}, by cutting $n<N$ ($N=n+\ell$) propagators, and hence unify all $\FFint_n^{(\ell)}$ with identical $n+\ell$ in a single object:
\begin{equation}
\begin{aligned}
    \FFint^{(\ell)}_n = \left(\sum_{\text{planar cut}}\mathrm{Cut}_n\right) \mathcal{I}_{N=n+\ell}\ ,\quad \text{where}\quad \mathcal{I}_{N} = \sum_{i}c^{(i)}_{N}\mathfrak{M}^{(i)}_{N}.\\
\end{aligned}
\end{equation}
Here by ``planar cut'' we refer to a cut that preserves the planarity of form factors at both sides. And the cutting procedure is equivalent to taking a periodic light-like limit on the integrand. We will spell out all these in the following sections.

\paragraph{Factorization properties}  Gauge theory amplitudes exhibit universal structures in the infrared sensitive regions \cite{Coleman:1965xm,Korchemskaya:1994qp,Korchemsky:1991zp,Kidonakis:1997gm,Collins:1988ig,Dixon:2008gr,Aybat:2006mz,Feige:2014wja}, which pose powerful constraints on the end-point behaviors of the integrand or the integrated functions. These properties, known as {\it hard-soft-collinear} factorization, naturally generalize to matrix element of local operators $F_{n}(\cO) $ \cite{Feige:2014wja}. 
The general statements of factorization refer to: universality in (multi-)collinear splitting, soft-gluon emissions, and the structures of virtual infrared divergences. These factorization properties can be formulated in terms of the form factor, which we translate into bootstrap equations imposed on the tree-level form factor squared, where soft limit is also translated into bootstrap equations on loop integrands. 

\subparagraph{Multi-collinear limit.}
In $\cN=4$ super Yang-Mills theory, scattering amplitudes and form factors are functions of on-shell kinematic degrees of freedom, namely on-shell momentum $\{ p_i= \lambda_i \tilde{\lambda}_i \} $ as well as super momentum $\{ q_i = \eta_i \lambda_i\} $.  In the limit where $N$ external particles go collinear, 
\begin{align}
  \lambda_i \rightarrow \sqrt{\omega_i}\,  \lambda_P \,,  \quad  \eta_i \rightarrow \sqrt{\omega_i} \, \eta_P \,,       \quad  \sum_{i} \omega_i = 1\,, \quad  i = 1, \ldots N,  
\end{align} 
the super-amplitude factorizes into a $1 \rightarrow N$ splitting amplitudes times lower point amplitudes \cite{Bern:1995ix,Kosower:1999xi,Bern:2004cz}:
\begin{align}
 A_{N+m}(1,2  \cdots, N+3)    \rightarrow  {\rm Split}_{1 \rightarrow N}  \times A_m(P, N+1, \cdots, N+m),
\end{align}
where ${\rm Split}_{1 \rightarrow N}$ is independent of the underlying lower-point scattering process.  The same argument holds for the super form factor. Hence the collinear limit of $(N+1)$-point squared form factor and 
$(N+3)$-point squared amplitudes are the same. Both are governed by the $1 \rightarrow N$ splitting function
\begin{align}
\lim_{1 \parallel \ldots \parallel N} \frac{ |A_{N+3}|^2 }{|A_4|^2}=  \lim_{1 \parallel  \ldots \parallel N}   \frac{\FFint^{(0)}_{N+1} }{\FFint^{(0)}_{2} }= |{\rm Split}_{1 \rightarrow N}|^2,
\end{align}
 where $|A_4|^2$ and $\FFint^{(0)}_2$ can be normalized to 1. 
 
The $N$-point splitting function is a scalar function of the kinematics in the collinear limit. Given the tree-level results available up to $N=11$ \cite{He:2024hbb}, a set of constraints can be imposed on the tree $\FFint^{(0)}_{N{+}1}$ in the limit where $p_1 \parallel \cdots \parallel p_{N}$.  In practice it is convenient for us to adopt spinor parametrization,  
\begin{align}
 | i\rangle =   \sqrt{\omega_i}  \, \big( |1\rangle  + z_i\, |N{+}1 \rangle \big) , \quad    | i] =   \sqrt{\omega_i}  \, \big( |1] + \bar{z}_i\, |N{+}1] \big),  \quad   i \in [2, N] .
\end{align} 
With this kinematic setup, we have 
\begin{align}
s_{i,N{+}1}/s_{1,N{+}1} = \omega_i , \quad s_{1i}/s_{1,N{+}1}= \omega_{i}|z_{i}|^2 ,  \quad s_{ij}/s_{1,N{+}1}= \omega_{i}\omega_{j} |z_{ij}|^2 , \quad i ,j \in [2, N] .
\end{align}
In the multi-collinear limit we send all the $z_i$'s to zero by taking  $(z_i, \bar{z}_i)  \rightarrow \epsilon (z_i,\bar z_i) $ and expand around $\epsilon =0$. 
$\FFint_{N{+}1}^{(0)}$ (modulo MHV) should agree with the splitting function $\mathcal{G}_{N} $ as given in \cite{He:2024hbb}, \textit{i.e.}
\begin{align}\label{eq:splittingfunc}
\lim_{\epsilon \rightarrow 0} \, \frac{\FFint^{(0)}_{N{+}1}}{ 2\, \FFint^{(0)}_{N{+}1,0}}   = \mathcal{G}_{N}, 
\end{align} 
once we set $x_{1} =1, z_1 = \bar{z}_1 = 0$ in $\mathcal{G}_{N}$ and identify $ x_i$ with $ \omega_i $,  for all  $i \in [2,  N]$. 

\subparagraph{Soft limit.}
We will proceed with a discussion on real soft emission from the tree-level form factors.  
Let us consider the $(N{+}1)$-point form factor in the limit where a single particle (e.g $p_{N{+}1}$) goes soft, the dominant contribution only comes from soft gluon emission with either positive or negative helicity. Hence we can write down a soft factorization formula, 
\begin{align}
\FFint^{(0)}_{N{+}1,k}  \rightarrow  \mathcal{S}^{(0)}_{N{+}1} \, \FFint^{(0)}_{N,k}  +  \mathcal{S}^{(0)}_{N{+}1} \, \FFint^{(0)}_{N,k-1} , \quad \FFint^{(0)}_{N{+}1,0}  \rightarrow  \mathcal{S}^{(0)}_{N{+}1} \, \FFint^{(0)}_{N,0}, 
\end{align}
where $\mathcal{S}^{(0)}_{N{+}1} = \frac{s_{N,1}}{s_{N,N{+}1} s_{N{+}1,1} }$ is the universal tree-level soft-gluon current \cite{Catani:2000pi}.  Taking the sum over the helicities on both sides, we observe that the $(N{+}1)$-point FF squared (modulo MHV)  reduces to $2$ times the $N$-point result
\begin{align}\label{eq:soft}
\lim_{p_{N{+}1} \rightarrow 0} \frac{\FFint^{(0)}_{N{+}1}}{ 2\, \FFint^{(0)}_{N{+}1,0}} = 2 \times \frac{\FFint^{(0)}_{N}}{ 2 \,\FFint^{(0)}_{N,0}} \,. 
\end{align}
Similar to the case for amplitude squared, it turns out that one can also impose exactly the same soft limit on loop integrands, $\FFint_n^{(\ell)}$ just as the tree-level case, which will be needed as an additional physical constraint starting $N=n+\ell=6$. 

\begin{figure}[htbp]
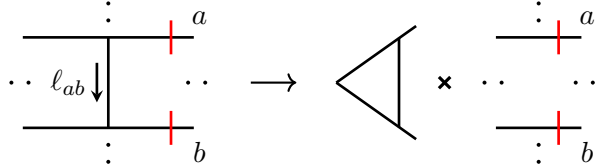

 \centering
  \softpinch
   \caption{Leading behaviour of a cut diagram in the one-loop virtual soft limit.}\label{fig:softpinch}
\end{figure}

Although not explicitly used here, similar observations can be made on limits of loop integrands with loop momenta becoming soft, where the universal structure of infrared divergences pose strong constraints. 
Let us consider the soft limit of a master diagram $\mathfrak{M}^{(i)}_{N}$  where loop momentum flowing through a single propagator goes soft. 
By power counting,  the diagram only gets enhanced if the soft propagator $l_{ab}$ is connected to two on-shell external legs $a$ and $b$,  as depicted in Fig.~\ref{fig:softpinch}.  
Suppose all such propagators in $\mathfrak{M}^{(i)}_{N}$  are collected into a set $\cL$. 
In the limit where any  $ l \in \cL $ goes soft,  the master diagram factorizes into the product of a one-loop triangle times a lower-loop diagram with propagator $l$ erased.  

The one-loop triangle contains soft-collinear divergences responsible for the $\epsilon$-IR poles upon integration. Meanwhile, the structure of IR poles in $\FFint_N$ is determined by the universal cusp anomalous dimension, which is a sum of dipoles at one loop level \cite{Becher:2009cu,Gardi:2009qi,Catani:1996jh}.  As a consequence,  the integrands must factorize and exponentiate in terms of the so-called soft-gluon {\it web} \cite{Gatheral:1983cz,Frenkel:1984pz,Sterman:1981jc,Gardi:2011wa,Gardi:2011yz} . Hence $\FFint_{n}^{(\ell)} $ in the one-loop soft limit are subject to the following set of constraints
\begin{align}
\sum_{ l \in \mathcal{L}} \,  \lim_{l \rightarrow 0} \FFint_{n}^{(\ell)} =\left[  \sum_{i=1}^n  V_{i,i+1}  (l_{i,i+1})  \right] \times   \FFint_{n}^{(\ell -1)} 
\end{align}
where $V_{a,b} $ is the one-loop web function, 
\begin{align}
 V_{a,b}  (l) := \frac{p_a \cdot p_b}{ (p_a \cdot l) \,  l^2 \,  (p_b \cdot l)}  \,. 
\end{align} 
 which is homogeneous in terms of the on-shell hard momenta $p_a$ and $p_b$. 

\section{Bootstrap: Master diagrams, Ansatz and Physical constraints}\label{sec:bootstrap}

In this section, we present our bootstrap program for the form factor squared, which includes: (1) how to construct the ansatz including the topologies of master diagrams and their numerators, and (2) how to fix all coefficients using physical constraints such as soft and multi-collinear limits. We will first outline the general method, and then move to detailed discussions for the warm-up cases $N=3,4$ and more non-trivial cases with new results for $N=5$ and $N=6$.

\begin{figure}[htpb]
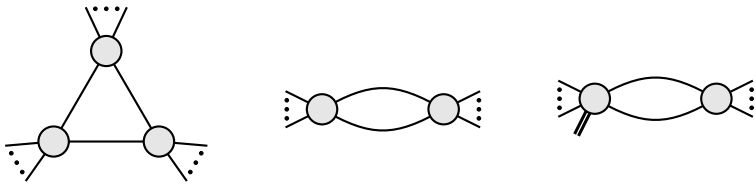

    \centering
    \scalebox{.8}{\trisubdiagram}
    \qquad
    \raisebox{1.5em}{\scalebox{.8}{\bubsubdiagram}}
    \qquad
    \raisebox{1.5em}{\scalebox{.8}{\bubsubdiagramFF}}
    \caption{The forbidden sub-diagrams.}
    \label{fig:trisubdiagram}
\end{figure}

\begin{figure}[htbp]
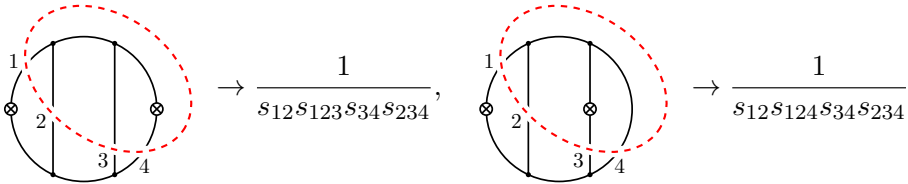

    \centering
    $$\vcenter{\hbox{\scalebox{.8}{\embedegA}}} \rightarrow \frac{1}{s_{12}s_{123}s_{34}s_{234}}, \quad \vcenter{\hbox{\scalebox{.8}{\embedegB}}} \rightarrow \frac{1}{s_{12}s_{124}s_{34}s_{234}}$$
    \caption{Two different embeddings of the same planar diagram.}
    \label{fig:embedding_eg}
\end{figure}

To write down the ansatz for the $N$-point master diagrams $\{\mathfrak{M}_N^{(1)}, \mathfrak{M}_N^{(2)},\ldots \}$, we should first list all of its possible top topologies. These top topologies are trivalent $(N-1)$-loop planar 1PI diagrams with two operators $\cO$'s inserted, where ``planar'' should be understood in the sense that operators are considered as insertion points instead of external legs\footnote{A planar diagram in our sense, could be non-planar when operators are considered as external legs.}. From $\cN=4$ cancellation, we see that any loop without $\cO$ insertion should have at least four interaction vertices~\cite{Bern:1994zx}, resulting in the well-known ``no-triangle/bubble'' rule which forbids any topology with triangle/bubble sub-diagram. For loops with operator $\cO$ insertion, we forbid the bubble sub-diagram~\cite{Boels:2012ew,Lin:2021qol,Lin:2021lqo}. The forbidden sub-diagrams are shown in Fig.~\ref{fig:trisubdiagram}. For $N=3,4,5,6$, we have $1,2,4,13$ top topologies\footnote{Actually there are $14$ top topologies for $N=6$, but one of them is excluded since it can only be obtained by adding ``rung" to lower-point topology with triangle sub-diagram. We will explain this in detail in Sec.~\ref{sec:n=6}.}, respectively, as shown in the first row of Table~\ref{tab:statAnsatz}. An important subtlety pointed out in~\cite{Bourjaily:2011hi} is that for a gauge theory like $\cN=4$ SYM, we should consider \textit{plane} diagrams (as opposed to just the above {\it planar} diagrams), because they have different color index flow. Recall that a plane diagram is a planar diagram with a specific plane embedding, and a planar diagram may have more than one inequivalent embedding, as shown in Fig.~\ref{fig:embedding_eg}. The number of plane diagrams with top topology for each $N$ is also shown in the first row of Table~\ref{tab:statAnsatz}.

With all these top-topology plane diagrams given, we then dress each diagram with possible numerators to obtain the ansatz. A crude method is to list all the monomials of the ISPs for each diagram according to dimension analysis, which always results in an ansatz with rather redundant parameters~\footnote{Note that the counting here depends on our particular choice of ISPs for each topology. Since we will focus on a much better method to write the ansatz, the ISPs will not be shown explicitly.}, as shown in the second row of Table~\ref{tab:statAnsatz}. To write down a more economic ansatz for master diagrams, power-counting of loop momentum and ``rung rule'' are applied and significantly eliminate the redundant parameters, as shown in the third row of Table~\ref{tab:statAnsatz}. Details of power-counting and ``rung rule'' will be explicitly explained in the following subsections.

Having the general ansatz as a linear combination of all $N$-point master diagrams at hand, $c_N^{(i)}$'s of $\sum_i c_N^{(i)} \mathfrak{M}_N^{(i)}$, we can cut them into the ansatz for $\FFint_n^{(\ell)}$ for different $\ell=N-n$ and impose physical constraints to determine all the coefficients $c_N^{(i)}$. The cutting should preserve the planarity of both sides, that is to say, cut propagators should be labeled in cyclic order, as shown in Fig.~\ref{fig:embedding_eg}. Note that different embeddings will result in cyclically different labeling of the cut propagators, thus producing cyclically different terms. For example, the two different embeddings in Fig.~\ref{fig:embedding_eg} are related by a non-cyclic permutation $(34)$, and give $\frac{1}{s_{12}s_{123}s_{34}s_{234}}$ and $\frac{1}{s_{12}s_{124}s_{34}s_{234}}$ respectively. With the ansatz given, we use the physical constraints, \textit{e.g.} Eq.\eqref{eq:splittingfunc} and/or Eq.\eqref{eq:soft} to determine $c_N^{(i)}$'s.

\begin{table}[ht]
    \centering
    \begin{tblr}{rowspec={|[1pt]Q[c,m]|Q[c,m]Q[c,m]Q[c,m]Q[c,m]Q[c,m]Q[c,m]|[1pt]},colspec={Q[c,m]|Q[c,m]Q[c,m]Q[c,m]Q[c,m]Q[c,m]}}
      $N$  & 3     & 4     & 5     & 6 \\
    \# of planar/plane diagrams &   1/1    &    2/3   &    4/6   &  13/21 \\
    \# of para. (ISP) &    -   &    2/3   &    49/69   &  1448/2118  \\
    \# of para. (rung rule) &    1   &    2/3   &   9/11   &  120/163 \\
    multi-collinear &    0   &   0    &    0   &  1 \\
    tree soft &    0   &   0    &    0   & 1 \\
    loop soft &   0    &    0   &     0  & 0 \\
    \end{tblr}
    \caption{Statistics of ansatz.}
    \label{tab:statAnsatz}
\end{table}

\begin{figure}[htbp]
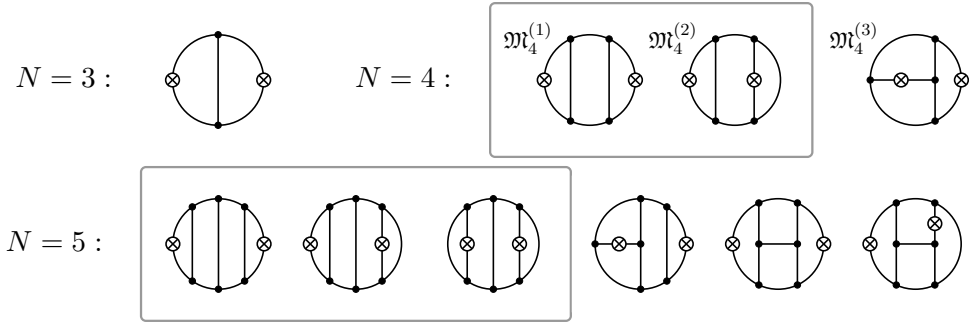

    \centering
    $N=3: \quad \vcenter{\hbox{\scalebox{.6}{\masterthree}}}$
    \qquad 
    $N=4: \quad \tcboxmath[colback=white,colframe=black!40!white,size=small,left=0mm,right=0mm]{
    \,\raisebox{15pt}{\scalebox{0.8}{$\mathfrak{M}_4^{(1)}$}}\hspace{-2.5ex}\vcenter{\hbox{\scalebox{.6}{\masterfourA}}}
    \!\!\raisebox{15pt}{\scalebox{0.8}{$\mathfrak{M}_4^{(2)}$}}\hspace{-2.5ex}\vcenter{\hbox{\scalebox{.6}{\masterfourB}}}}
    \;\;\raisebox{15pt}{\scalebox{0.8}{$\mathfrak{M}_4^{(3)}$}}\hspace{-2.5ex}\vcenter{\hbox{\scalebox{.6}{\masterfourC}}}$\\[2pt]
    $N=5: \quad \tcboxmath[colback=white,colframe=black!40!white,size=small,left=0mm,right=0mm]{\vcenter{\hbox{\scalebox{.6}{\masterfiveA}}}
    \vcenter{\hbox{\scalebox{.6}{\masterfiveB}}}
    \vcenter{\hbox{\scalebox{.6}{\masterfiveC}}}}
    \vcenter{\hbox{\scalebox{.6}{\masterfiveD}}}
    \vcenter{\hbox{\scalebox{.6}{\masterfiveE}}}
    \vcenter{\hbox{\scalebox{.6}{\masterfiveF}}}$
    \caption{Plane diagrams of top topology for $N=3,4,5$, where top topologies for $N=4$ coincide with the master diagrams. Diagrams in the same grey box correspond to the same planar diagram.}
    \label{fig:345pt_all}
\end{figure}

\subsection{\texorpdfstring{$N=3, 4$}{N=3,4}}\label{sec:3pt4pt}
Let us start with the trivial example of $N=3$, where we only have 1 plane diagram (see the first diagram in Fig.~\ref{fig:345pt_all}). Cutting 3 propagators and summing over cyclic permutations of the cut propagators will give:
\begin{equation}\label{eq:3ptcut}
    \frac{1}{4} \times \hspace{-1em} \sum_{\text{cyc}\{1,2,3\}}\left(  \vcenter{\hbox{\scalebox{.6}{\cutthreeA}}} + \vcenter{\hbox{\scalebox{.6}{\cutthreeB}}} \right) = \frac{q^2}{s_{12}s_{23}s_{31}},
\end{equation}
where we have included a symmetry factor of $1/4$ to make the R.H.S. have unit coefficient. We will always do so for higher points. And note that the symmetry factor is just one over the group order of the automorphism group of the plane diagram. In this case, according to dimension analysis, we should have included a numerator of mass dimension ``$-1$", which manifests itself as $1/q^2$ from Eq.\eqref{eq:3ptcut}. Noting that $\FFint_3^{(0)} = \frac{2}{s_{12}s_{23}s_{31}}$, we then have $c_3^{(1)}=2$.

Let us move to the next case, $N=4$, where we have three plane diagrams, see Fig.~\ref{fig:345pt_all}. The first two diagrams correspond to the same planar diagram, as explained in Fig.~\ref{fig:embedding_eg}. In this case, the numerator goes like $s^0$ (constant), so we can simply identify the plane diagrams as $\{\mathfrak{M}_4^{(1)},\mathfrak{M}_4^{(2)},\mathfrak{M}_4^{(3)}\}$. Cutting 4 propagators of them in a cyclic order, we have:
\begin{equation}\label{eq:N4cuts}
\begin{aligned}
  &\mathfrak{M}_4^{(1)} \rightarrow \frac{1}{4} \times \hspace{-1em} \sum_{\text{cyc}\{1,2,3,4\}} \left( \vcenter{\hbox{\scalebox{.6}{\cutfourAone}}} + \vcenter{\hbox{\scalebox{.6}{\cutfourAtwo}}} \right)
  =\frac{1}{s_{12}s_{34}s_{123}s_{341}} + \text{cyc.}\\
  &\mathfrak{M}_4^{(2)} \rightarrow \frac{1}{4} \times \hspace{-1em} \sum_{\text{cyc}\{1,2,3,4\}} \left( \vcenter{\hbox{\scalebox{.6}{\cutfourBone}}} + \vcenter{\hbox{\scalebox{.6}{\cutfourBtwo}}} \right)
  =\frac{1}{s_{12}s_{34}s_{412}s_{341}} + \text{cyc.}\\
   &\mathfrak{M}_4^{(3)} \rightarrow \frac{1}{8} \times \hspace{-1em} \sum_{\text{cyc}\{1,2,3,4\}} \left( \vcenter{\hbox{\scalebox{.6}{\cutfourCone}}}\hspace{-0.5em} + \hspace{-0.5em} \vcenter{\hbox{\scalebox{.6}{\cutfourCtwo}}} \hspace{-0.5em} + \hspace{-0.5em} \vcenter{\hbox{\scalebox{.6}{\cutfourCthree}}} \hspace{-0.5em} + \hspace{-0.6em} \vcenter{\hbox{\scalebox{.6}{\cutfourCfour}}} \hspace{-0.5em} + \hspace{-0.6em} \vcenter{\hbox{\scalebox{.6}{\cutfourCfive}}} \right) \\
   & \hspace{3.2em} =\frac{1}{s_{12}s_{23}s_{34}s_{41}} + 
    \frac{1}{s_{12}s_{23}s_{412}s_{234}} + \text{cyc.}\\
\end{aligned}
\end{equation}
where $1/4$ and $1/8$ are the symmetry factors of each diagram, and ``cyc.'' means all cyclic inequivalent images of previous terms. We then have an ansatz for $\FFint_4^{(0)}$:
\begin{equation}\label{eq:N4ansatz}
\begin{aligned}
    & \FFint_4^{(0)} = c_4^{(1)} \left(\frac{1}{s_{12}s_{34}s_{123}s_{341}} + \text{cyc.}\right)
    + c_4^{(2)} \left( \frac{1}{s_{12}s_{34}s_{412}s_{341}} + \text{cyc.} \right) \\
    & \hspace{9.1em} + c_4^{(3)} \left( \frac{1}{s_{12}s_{23}s_{34}s_{41}} + 
    \frac{1}{s_{12}s_{23}s_{412}s_{234}} + \text{cyc.} \right).
\end{aligned}
\end{equation}
Imposing multi-collinear limit on both sides, we obtain $c_4^{(1)}=c_4^{(2)}=c_4^{(3)}=2$. Note that in both of these ``warm-up" cases, no non-trivial numerators are needed and one can fix these coefficients even by {\it e.g.} comparing with the square of BCFW results, but we will see that it becomes much more non-trivial for $N\geq 5$.

\subsection{\texorpdfstring{$N=5$}{N=5}}
Things start to become interesting for $N=5$. There are 6 plane diagrams shown in the second row of Fig.~\ref{fig:345pt_all} (the first three correspond to the same planar diagram, thus we have 4 top topologies in total). The numerators should be $s^1$, and naively they can be chosen from 11 propagators (which lead to sub-topologies), 3 ISPs and $q^2$, and give $ 6 \times 15 = 90$ master diagrams in total. After considering the symmetry of the master diagrams, we are left with $ 69$ master diagrams. After cutting $5$ propagators, we will have an ansatz for $\FFint_5^{(0)}$. We found that the $69$ terms in the $\FFint_5^{(0)}$ ansatz are not linearly independent, and some of them do not contribute to the multi-collinear limit or (tree-level) soft limit, making the crude ansatz too large to be determined by tree-level physical constraints.

\begin{figure}[htbp]
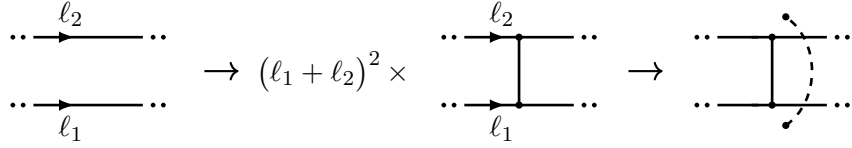

    \centering
    {\rungrule}
    \caption{Rung rule. When adding a ``rung" between two propagators with momenta $\ell_1$ and $\ell_2$, we should write a numerator $(\ell_1+\ell_2)^2$, which is further denoted as a dashed line in the diagram.}
\label{fig:rungrule}
\end{figure}

\begin{figure}[htbp]
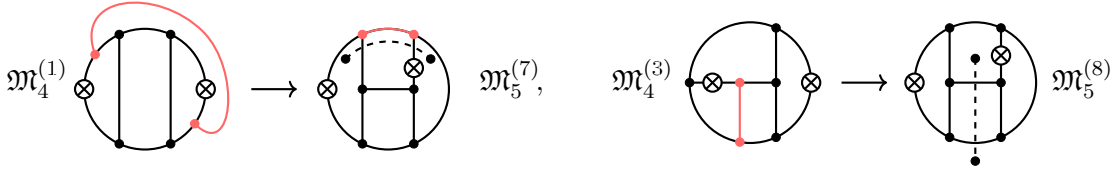

    \centering
    $\mathfrak{M}_4^{(1)} \hspace{-0.5em} \vcenter{\hbox{\scalebox{.8}{\rungruleEgA}}} \mathfrak{M}_5^{(7)} \hspace{-0.5em}$ , \qquad
    $\mathfrak{M}_4^{(3)} \hspace{-0.5em} \vcenter{\hbox{\scalebox{.8}{\rungruleEgB}}} \hspace{-0.5em} \mathfrak{M}_5^{(8)}$ 
    \caption{We can add a rung (the red lines) to a master diagram of $N=4$ and get a master diagram of $N=5$, where a dashed line denotes a numerator. Adding rung to different diagrams may result in same topology but with different numerator.}
    \label{fig:rungruleEG}
\end{figure}

\begin{figure}[htbp]
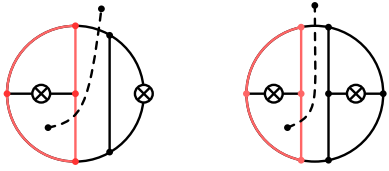

    \centering
    \scalebox{.9}{\vanishingrungrule}
    \caption{We discard these two master diagram generated by rung rule, since the ISP devastates the power-counting of the 4-cycle highlighted in red.}
    \label{fig:5pt_exclude}
\end{figure}

\begin{figure}[htbp]
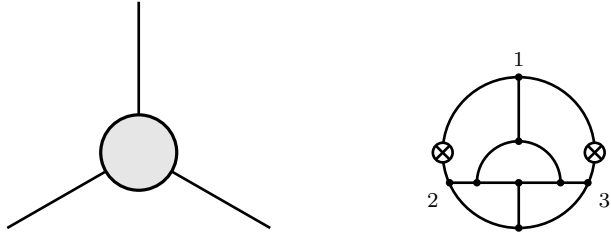

    \centering
    \vanishingsubdiagram
    \caption{A general 3-point sub-diagram is shown on the left. And the $N=6$ top topology containing 3-point sub-diagram is shown on the right.}
\label{fig:3ptsubdiagram}
\end{figure}

However, by smartly generating the ansatz with the help of correct power-counting and the ``rung rule"~\cite{Bern:1997nh}, we can obtain a more compact ansatz that can be determined by multi-collinear limit of $\FFint_5^{(0)}$. To be explicit, the rung rule states that a ``rung'' added between two propagators with momenta $\ell_1$ and $\ell_2$ contributes a factor of $(\ell_1+\ell_2)^2$ for the numerator, which is further denoted by a dashed line on the diagram, as shown in Fig.~\ref{fig:rungrule}. Most of the master diagrams, including all of those with top topology, can be generated by adding rungs to lower-point master diagrams. Two examples are shown in Fig.~\ref{fig:rungruleEG}. However, not all the diagrams generated by rung rule have the correct power counting property, \textit{e.g.} the diagram shown on the left of Fig.~\ref{fig:5pt_exclude}, where the ISP devastates the power-counting of the 4-cycle highlighted in red. We will discard this master diagram in our ansatz. It is also noteworthy that, if we suppose that rung rule is able to give the complete diagrams with top topology, any topology containing a non-trivial 3-point sub-diagram will be excluded \footnote{Also recall that there is no higher-loop correction to the 3-point amplitudes in $\cN=4$ SYM. And generalized unitarity will prevent 3-point sub-diagrams~\cite{Engelund:2012re}.}, as shown in the left of Fig.~\ref{fig:3ptsubdiagram}, since they can only arise by adding rungs to lower-point topologies with sub-triangles.

\begin{figure}[htbp]
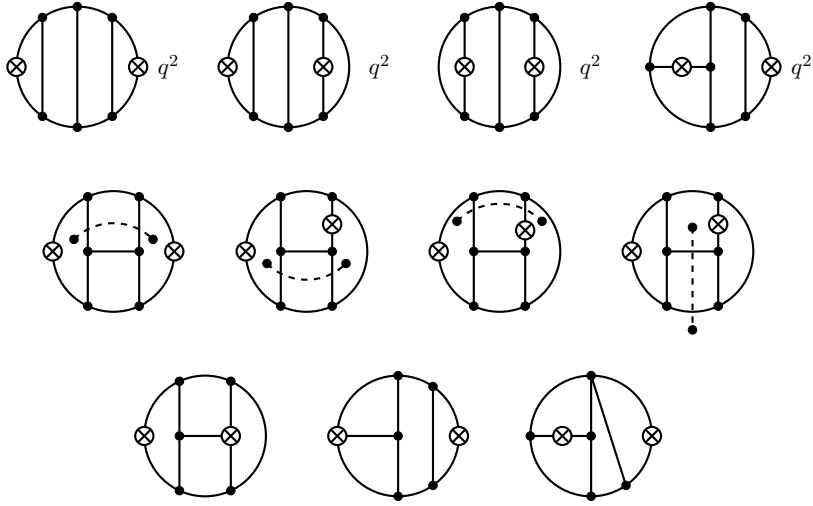

    \centering
    \scalebox{.8}{\masterfiveOne}
    \scalebox{.8}{\masterfiveTwo}
    \scalebox{.8}{\masterfiveThree}
    \scalebox{.8}{\masterfiveFour}\\
    \scalebox{.8}{\masterfiveFive}
    \scalebox{.8}{\masterfiveSix}
    \scalebox{.8}{\masterfiveSeven}
    \scalebox{.8}{\masterfiveEight}\\
    \scalebox{.8}{\masterfiveNine}
    \scalebox{.8}{\masterfiveTen}
    \scalebox{.8}{\masterfiveEleven}
    \caption{$N=5$ master diagrams. We denote them as $\{\mathfrak{M}_5^{(1)},\ldots,\mathfrak{M}_5^{(11)}\}$.}
    \label{fig:5ptall}
\end{figure}

For the master diagrams with sub-topologies, we use power-counting to further constrain the ansatz: for any propagator contained in an $m$-cycle loop, the maximal power of its momentum square in the numerator is $m-4$ (or $m-3$, if an operator $\cO$ is inserted in the loop). Consequently, we only need to shrink propagators contained in an $m$-cycle with $m>4$ (or $m>3$ if it has an operator $\cO$ inserted). We list all the master diagrams in Fig.~\ref{fig:5ptall}, and denote them as $\{\mathfrak{M}_5^{(1)},\ldots,\mathfrak{M}_5^{(11)}\}$ respectively. 

Cutting $5$ legs of the master diagrams, we will have two types of rational functions. The first type, stemming from cutting the top-topologies, has nontrivial numerators, taking $\mathfrak{M}_5^{(5)}$ for example:
\begin{equation}
    \vcenter{\hbox{\scalebox{.8}{\cutfiveA}}} = \frac{s_{234}}{s_{12}s_{34}s_{1234} s_{23}s_{45}s_{2345}}.
\end{equation}
And the second type, obtained when cutting a diagram with sub-topology, has numerator ``1", taking $\mathfrak{M}_5^{(9)}$ for example:
\begin{equation}
    \vcenter{\hbox{\scalebox{.8}{\cutfiveB}}} = \frac{1}{s_{12}s_{34}s_{1234} s_{45}s_{451}}.
\end{equation}
After cutting all the diagrams, we have an ansatz that is linearly independent and all the terms contribute to multi-collinear limit. By imposing multi-collinear limit to it, we determine the coefficients of each diagram as:
\begin{equation}
    \{c_5^{(1)},\ldots,c_5^{(11)}\}=\{2,2,2,2,2,2,0,2,-2,0,-2\}.
\end{equation}
From this result, we can cut $n<5$ propagators to get $\FFint_n^{(N-n)}$, especially we can extract the integrand of $3$-loop Sudakov FF as will be explained in Sec.~\ref{sec:loopintegrands}.

\subsection{\texorpdfstring{$N=6$}{N=6}}\label{sec:n=6}

\begin{figure}[htbp]
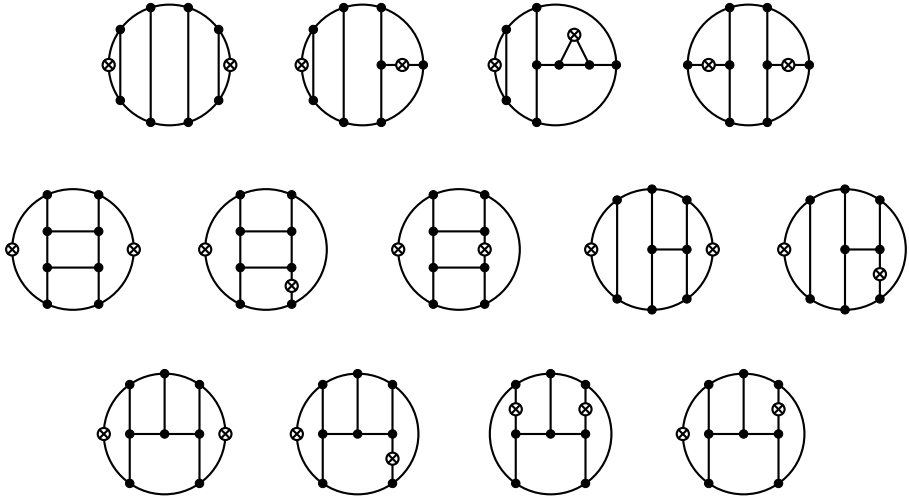

    \centering
    \scalebox{.8}{\mastersixOne}
    \scalebox{.8}{\mastersixTwo}
    \scalebox{.8}{\mastersixThree}
    \scalebox{.8}{\mastersixFour}\\
    \scalebox{.8}{\mastersixFive}
    \scalebox{.8}{\mastersixSix}
    \scalebox{.8}{\mastersixSeven}
    \scalebox{.8}{\mastersixEight}
    \scalebox{.8}{\mastersixNine}\\
    \scalebox{.8}{\mastersixTen}
    \scalebox{.8}{\mastersixEleven}
    \scalebox{.8}{\mastersixTwelve}
    \scalebox{.8}{\mastersixThirteen}
    \caption{Top topologies of $N=6$.}
    \label{fig:6pttop}
\end{figure}

For $N=6$, we have 13 top topologies (21 plane diagrams) to consider, as shown in Fig.~\ref{fig:6pttop}, where we have already discarded 1 topology shown in the right of Fig.~\ref{fig:3ptsubdiagram}, for it contains a $3$-point sub-diagram (with vertices marked by $1$, $2$ and $3$). In this subsection, we show how to obtain an ansatz with 163 parameters in detail.

We first apply rung rule to the master diagrams with top topologies of $N=5$. All of the $N=6$ top topologies can be generated in this procedure. But for the last topology on the first row of Fig.~\ref{fig:6pttop}, the corresponding master diagram obtained by rung rule (shown on the right of Fig.~\ref{fig:5pt_exclude}) is eliminated by power-counting. For this top topology, we choose the numerator to be $(q^2)^2$, because it is the only mass-dimension-two numerator that doesn't devastate the power-counting property. Adding it to the ansatz, we will have $40$ master diagrams for the $N=6$ top topologies. To obtain the sub-topologies, we start from the $40$ master diagrams of the top topologies. Note that each of them has two numerators, we first choose one of the numerators and consider all possible ways of pinching a propagator that preserves the power-counting property. This will give us a master diagram of sub-topology. Doing this for all the $40$ master diagrams and for both numerators of each diagram, we obtain $100$ master diagrams of sub-topology (and there are $20$ different sub-topologies in total). At last, for the sub-sub-topology, we only need to consider all possibilities of pinching two propagators of one top topology using power-counting constraint. We then have $23$ master diagrams of sub-sub-topology. In total, we have $40+100+23=163$ terms in our ansatz.

After cutting $6$ propagators of this ansatz, we found that they are not linearly independent, and there is one term that does not contribute to the multi-collinear limit. And this term does not contribute to the tree-level soft limit either. So we need to determine its coefficient using constraints from loop-level soft limit. Consider cutting 3 propagators, and take the soft limit, which will give the $3$-loop Sudakov FF squared. Comparing with the result obtained by cutting 2 propagators of $N=5$, the undetermined coefficient turns out to be $0$. We thus determine all the parameters. Note that there are still free parameters remained, due to the linear relation among the master diagrams. We can choose the value of these parameters at will, and finally obtain $59$ contributing master diagrams. The corresponding integrand is provided in the ancillary file. We have also checked our results of $\FFint_{6}^{(0)}$ by numerically comparing to a direct calculation by BCFW recursion, which we review in Appendix.~\ref{sec:bcfw}.

\section{Results: Trees and Loop Integrands}

Since we have obtained the coefficients of the master diagrams with $N=3,4,5,6$, we can cut $n<N$ propagators to get $\FFint_n^{(\ell)}$. To unify the procedure of cutting master diagrams into form factor squared, it will be helpful to introduce the periodic dual space description for master diagram integrand (equivalent class) $\mathcal{I}_N$, since the cutting procedure is simply taking a periodic light-like limit of this integrand. In this section, we will first describe how to write the integrand $\mathcal{I}_N$ in the periodic dual space, and then show our result of $\FFint_n^{(\ell)}$ for different $n$ and $\ell$ in both momentum space and periodic dual space. We also provide an ancillary file recording the results in periodic dual space.

\subsection{Periodic dual space}\label{sec:dualcoordinates}
\begin{figure}[htbp]
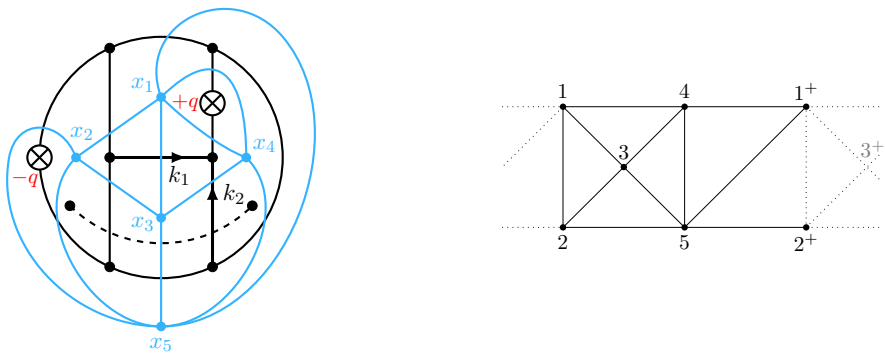

    \centering
    \begin{minipage}[c]{0.45\textwidth}
    \centering\scalebox{.8}{\dualEG}
    \end{minipage}
    \begin{minipage}[c]{0.45\textwidth}
    \centering\scalebox{.8}{\periodicdualEG}
    \end{minipage}
    \caption{$\mathfrak{M}_5^{(6)}$ and its dual graph (the blue graph).}
    \label{fig:dualcoordinates}
\end{figure}

To obtain the periodic dual space description of the integrand, we first consider the master diagram as a graph on a two-punctured sphere with the two operators considered as punctures, and draw the dual graph of the master diagram, as shown on the left of Fig.~\ref{fig:dualcoordinates}. Then we assign each dual point a coordinate $x_i$, so that the momentum flowing through a propagator between two adjacent regions $i,j$ in master diagram is given by the difference of the corresponding dual coordinates $x_{ij}:=x_i-x_j$, \textit{e.g.}  $k_1^2=x_{13}^2$, $k_2^2=x_{34}^2$ in Fig.~\ref{fig:dualcoordinates}. Note that a two-punctured sphere is topologically equivalent to a cylinder, we can view the dual graph as being on the cylinder, and ``unfold'' it to get the periodic dual graph, as shown on the right of Fig.~\ref{fig:dualcoordinates}. This enables us to assign the endpoint obtained by traversing a counterclockwise loop around puncture $\pm q$ starting from $x_i$ by $x_{i^\pm}:=x_i \pm q$, and thus distinguish between two propagators separated by an operator between two regions by denoting one as $x_{ij}$ and the other as $x_{ji^+}$. For example, the propagators below and above puncture $+q$ between $x_1$ and $x_4$ in Fig.~\ref{fig:dualcoordinates} are given by $x_{14}$ and $x_{41^+}$, respectively.

With the above definition of periodic dual coordinates, we can translate all inverse propagators into $x_{ij}^2$'s, therefore reduce the ambiguity of the integrand from arbitrary loop momentum shift to only periodic shift $x_i \to x_{i^{(+k_i)}}:=x_i+k_i\mspace{1mu} q$, and resolve the ambiguity of propagator labeling by summing over all permutations of the dual points divided by the symmetry factor. For example, the master diagram $\mathfrak{M}_5^{(6)}$ is now given by the following rational function, up to periodic shift at some of its points:
\begin{equation}\label{eq:seed}
    \mathfrak{M}_5^{(6)}=\frac{1}{2}\left(\frac{x_{24}^2}{x_{12}^2 x_{13}^2 x_{14}^2 x_{23}^2 x_{25}^2 x_{34}^2 x_{35}^2 x_{45}^2 x_{41^{+}}^2 x_{51^{+}}^2 x_{52^{+}}^2} +\text{all perms}\right).
\end{equation}
Here the term ``all permutations'' encompasses operations generated by permutations of the $N$ dual points and the reflection of $q$. This also includes the operations that reverse the $+q$ direction of the periodic dual graph, {\it e.g.} rotating the periodic dual graph of $\mathfrak{M}_5^{\raisemath{-2pt}{(6)}}$ by $180^\circ$ around $x_3$ (which is the only non-trivial automorphism of the periodic dual graph of $\mathfrak{M}_5^{\raisemath{-2pt}{(6)}}$, making its symmetry factor $2$). For more examples for the master diagrams and their corresponding (periodic) dual graphs with integrands, see Table~\ref{tab:periodic_dual_graph_4pt}.

\begin{figure}[htbp]
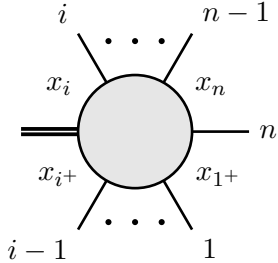

    \centering
    $\vcenter{\hbox{\scalebox{1}{\dualEXT}}}$
    \caption{The relation between external legs and periodic dual coordinates. The operator leg is inserted between external legs $i-1$ and $i$.}
    \label{fig:dualEXT}
\end{figure}

Similar to the story of $f$-graph, the $n$-point form factor squared is given by taking the light-like limit of the periodic $n$-cycle $(123\dots n1^+)$. However, due to the ambiguity of periodic shift, naively we should consider not only the light-like limit of $x_{12}^2,x_{23}^2,\dots,x_{n1^+}^2$, but also any of its equivalent shifted version, {\it e.g.} $x_{12^+}^2,x_{2^+3}^2,\dots,x_{n1^+}^2$ shifted by $x_2 \to x_{2^+}$. In order to obtain form factor squared by a single light-like limit instead of infinitely many equivalent ones, we will adopt the following prescription for integrand: any propagator $x_{ij}$ in the integrand should have $x_i,x_j$ within some common period, {\it e.g.} $x_{12},x_{41^+}$, and any periodically shifted version of the integrand involving $x_{ij}$ spanning over one period like $x_{12^+},x_{4^-1^+}$ is forbidden. This prescription excludes most of the ambiguity of periodic shift, leaving only a finite number (sometimes only one) of equivalent integrands for a master diagram. In particular, this prescription leaves $x_{12}^2,x_{23}^2,\dots,x_{n1^+}^2$ as the only contributing light-like limit, excluding any of its periodically shifted version like $x_{12^+}^2,x_{2^+3}^2,\dots,x_{n1^+}^2$. With the above prescription, the cut form factor squared integrand is simply given by
\begin{equation}\label{eq:LLlimit}
   \FFint_n^{(\ell=N-n)} = \lim_{\substack{x_{i,i+1}^2 \rightarrow 0,\\
    1 \leq i \leq n }}  \left[ \left( \prod_{1 \leq i \leq n} x_{i,i+1} \right) \, \mathcal{I}_N \right] ,
\end{equation}
where we should identify $x_{n+1} = x_{1^{+}}$ due to periodicity. This limit is equivalent to cutting $n$ propagators in a cyclic compatible way, \textit{i.e.} the external legs are cyclically labeled with $1 \ldots n$. Since the operator leg (denoted by double line) can appear between any pair of adjacent external legs, say $i-1$ and $i$, then the relation between external legs and periodic dual coordinates is illustrated in Fig.~\ref{fig:dualEXT}.

\begin{table}[ht]
    \centering
    \begin{tblr}{rowspec={|[1pt]Q[c,m]|Q[c,m]Q[c,m]Q[c,m]|[1pt]},colspec={Q[c,m]Q[c,m]Q[c,m]}}
         master diagram & periodic dual graph & integrand perm seed \\
         \raisebox{30pt}{$\mathfrak{M}_4^{(1)}$}\hspace{-5.5ex}{\begin{minipage}[c]{0.25\textwidth}\centering\scalebox{0.55}{\dualfourA} \end{minipage}} & {\begin{minipage}[c]{0.25\textwidth}\centering\scalebox{0.6}{\periodicdualfourA} \end{minipage}} & \scalebox{0.85}{$\displaystyle\frac{1}{4}\frac{1}{x_{12}^2 x_{13}^2 x_{14}^2 x_{23}^2 x_{21^+}^2 x_{34}^2 x_{31^+}^2 x_{41^+}^2}\;$} \\
         \raisebox{30pt}{$\mathfrak{M}_4^{(2)}$}\hspace{-5.5ex}{\begin{minipage}[c]{0.25\textwidth}\centering\scalebox{0.55}{\dualfourB} \end{minipage}} & {\begin{minipage}[c]{0.25\textwidth}\centering\scalebox{0.6}{\periodicdualfourB} \end{minipage}} & \scalebox{0.85}{$\displaystyle\frac{1}{4}\frac{1}{x_{12}^2 x_{13}^2 x_{14}^2 x_{24}^2 x_{34}^2 x_{31^+}^2 x_{41^+}^2 x_{42^+}^2}\;$} \\
         \raisebox{30pt}{$\mathfrak{M}_4^{(3)}$}\hspace{-5.5ex}{\begin{minipage}[c]{0.25\textwidth}\centering\scalebox{0.55}{\dualfourC} \end{minipage}} & {\begin{minipage}[c]{0.3\textwidth}\centering\scalebox{0.6}{\periodicdualfourC} \end{minipage}} & \scalebox{0.85}{$\displaystyle\frac{1}{8}\frac{1}{x_{12}^2 x_{13}^2 x_{23}^2 x_{24}^2 x_{34}^2 x_{31^+}^2 x_{41^+}^2 x_{42^+}^2}\;$} \\
    \end{tblr}
    \caption{$N=4$ master diagrams together with their (periodic) dual graphs and integrands.}
    \label{tab:periodic_dual_graph_4pt}
\end{table}

\subsection{Trees}
In this subsection, we show the result of $\FFint_N^{(0)}$, which can be used as the integrand of energy correlator.

For $N=3$, it is easy to see from Eq.~\eqref{eq:3ptcut} that $\FFint_3^{(0)} = \frac{2}{s_{12}s_{23}s_{31}}$ in momentum space. Then we will work with the dual coordinates. Firstly, we write down the integrand as
\begin{equation}
    \mathcal{I}_3 = \frac{2}{x_{12}^2 x_{13}^2 x_{23}^2 x_{21^{+}}^2 x_{31^{+}}^2 x_{11^{+}}^2} + \frac{2}{x_{23}^2 x_{21^{+}}^2 x_{31^{+}}^2 x_{32^{+}}^2 x_{12}^2 x_{22^{+}}^2} + \frac{2}{x_{31^{+}}^2 x_{32^{+}}^2 x_{12}^2 x_{13}^2 x_{23}^2 x_{33^{+}}^2}.
\end{equation}
Taking the light-like limit gives
\begin{equation}
    \FFint_3^{(0)} = \lim_{\substack{x_{i,i+1}^2 \rightarrow 0,\\
    1 \leq i \leq 3 }}  \left[ \left(x_{12}^2 x_{23}^2 x_{31^{+}}^2 \right) \, \mathcal{I}_3 \right] = \frac{2(x_{13}^2 + x_{21^{+}}^2 + x_{32^{+}}^2)}{q^2 x_{13}^2 x_{21^{+}}^2 x_{32^{+}}^2} = \frac{2}{ x_{13}^2 x_{21^{+}}^2 x_{32^{+}}^2},
\end{equation}
where we use the identity $(x_{13}^2 + x_{21^{+}}^2 + x_{32^{+}}^2)=q^2$. It is clearly the same as $\frac{2}{s_{12}s_{23}s_{31}}$.

Next we move on to $\FFint_4^{(0)}$. Recall from Eq.~\eqref{eq:N4ansatz}:
\begin{equation}\label{eq:4ptTreeMandel}
\begin{aligned}
    \FFint_4^{(0)} & = 2 \left(\frac{1}{s_{12}s_{23}s_{34}s_{41}} + 
    \frac{1}{s_{12}s_{23}s_{412}s_{234}} +\frac{1}{s_{12}s_{34}s_{123}s_{341}} + \frac{1}{s_{12}s_{34}s_{412}s_{341}} + \text{cyc.} \right)\\
    & = \frac{2}{s_{12}s_{23}s_{34}s_{41}} \left( 1 +\frac{s_{23}s_{41}}{s_{123}s_{341}} + \frac{s_{23}s_{41}}{s_{412}s_{341}} + \frac{s_{34}s_{41}}{s_{412}s_{234}} + \text{cyc.} \right),
\end{aligned}
\end{equation}
which is consistent with \cite{Yan:2022cye}. There are $4$ cyclic seeds in total for $N=4$, and each of them corresponds to one of the cuts of a master diagram, see Eq.~\eqref{eq:N4cuts}. This result is also straightforward if we work with dual coordinates \footnote{In Eq.~\eqref{eq:4ptTreeMandel} and \eqref{eq:4ptTreeDual}, we factor out the MHV part $\FFint_{4,0}^{(0)}$ and rewrite the result in terms of ``cross-ratios'' in the last equality. Comment is that the ``cross-ratio'' expression $\FFint_4^{(0)}/\FFint_{4,0}^{(0)}$ enjoys a special symmetry of ``directed dual conformal invariance'' (DDCI) in the limit of $q^2 \rightarrow 0$ \cite{Bern:2018oao,Chicherin:2018wes,Bianchi:2018rrj}. We have also checked that $\FFint_{5}^{(0)}/\FFint_{5,0}^{(0)}$ is DDCI, which we won't give more details, since it is beyond the scope of this paper.}:
\begin{equation}\label{eq:4ptTreeDual}
\begin{aligned}
    & \FFint_4^{(0)} = \lim_{\substack{x_{i,i+1}^2 \rightarrow 0,\\
    1 \leq i \leq 4 }}  \left[ \left(x_{12}^2 x_{23}^2 x_{34}^2 x_{41^{+}}^2 \right) \, \mathcal{I}_4 \right] \\
    & = 2 \left(\frac{1}{x_{13}^2 x_{24}^2 x_{31^{+}}^2 x_{42^{+}}^2}  + 
    \frac{1}{x_{13}^2 x_{24}^2 x_{43^{+}}^2 x_{21^{+}}^2} +\frac{1}{x_{13}^2 x_{31^{+}} x_{14}^2 x_{32^{+}}^2 } + \frac{1}{x_{13}^2 x_{31^{+}}^2 x_{43^{+}}^2 x_{32^{+}}^2} + \widetilde{\text{cyc.}} \right)\\
    & = \frac{2}{x_{13}^2 x_{24}^2 x_{31^{+}}^2 x_{42^{+}}^2} \left(1 + 
    \frac{x_{31^+}^2 x_{42^+}^2}{x_{43^+}^2 x_{21^+}^2} +\frac{x_{24}^2 x_{42^+}^2}{x_{14}^2 x_{32^+}^2} + \frac{x_{24}^2 x_{42^+}^2}{x_{43^+}^2 x_{32^+}^2} + \widetilde{\text{cyc.}} \right),
\end{aligned}
\end{equation}
where $\widetilde{\text{cyc.}}$ means the cyclic should be understood in the periodic sense. The integrand $\mathcal{I}_4$ can be obtained as explained in Sec.~\ref{sec:dualcoordinates}, and can be found in the ancillary file.

For $N=5$, we can also obtain the result from two perspectives, \textit{i.e.} cutting propagators of master diagrams in the momentum space as well as working with periodic dual coordinates. We have found $220$ terms appearing in $\FFint_5^{(0)}$, where we count the terms by different cuts of a master diagram. More explicitly, taking $\mathfrak{M}_5^{(5)}$ for example, it has $4$ cyclically inequivalent cuts:
\begin{equation}
\begin{aligned}
    & \vcenter{\hbox{\scalebox{.8}{\cutfiveA}}} = \frac{s_{234}}{s_{12}s_{34}s_{1234} s_{23}s_{45}s_{2345}}, & \hspace{-1em}
    \vcenter{\hbox{\scalebox{.8}{\cutfiveTreeA}}} = \frac{s_{234}}{s_{34}s_{345}s_{2345} s_{23}s_{123}s_{1234}},\\
    & \vcenter{\hbox{\scalebox{.8}{\cutfiveTreeB}}} \hspace{-1em} = \frac{s_{23}}{s_{34}s_{234}s_{2345} s_{12}s_{123}s_{45}}, & \hspace{-2em}\vcenter{\hbox{\scalebox{.8}{\cutfiveTreeC}}} = \frac{s_{34}}{s_{23}s_{234}s_{1234} s_{45}s_{345}s_{12}},
\end{aligned}
\end{equation}
which correspond to $4$ cyclic seeds. Note that the two terms on the second line are related by reflection. So if we consider dihedral seeds, the number will be $3$. For $N=4$, the cyclic and dihedral seeds are equivalent, so we didn't mention the difference in previous discussion. Counting every contributing master diagram in the same way, the numbers of cyclic (dihedral) seeds are $\{1,2,1,4,4,9,9,6,8\}$ ($\{1,1,1,2,3,6,6,3,4\}$) respectively for $\{ \mathfrak{M}_5^{(1)},\mathfrak{M}_5^{(2)},\ldots, \mathfrak{M}_5^{(6)}, \mathfrak{M}_5^{(8)},\mathfrak{M}_5^{(9)},\mathfrak{M}_5^{(11)} \}$. In total, there are $44$ cyclic seeds (or $27$ dihedral seeds) in the result. And we provide an ancillary file that contains the whole result of $\FFint_5^{(0)}$ in terms of dual coordinates. For $N=6$, the $59$ contributing master diagrams give $16703$ cyclic seeds ($8699$ dihedral seeds). We put all the results in the ancillary file.

\subsection{Loop integrands}\label{sec:loopintegrands}

In this subsection, we describe how to obtain $\FFint_{n}^{(\ell)}$ by cutting $n < N$ propagators, with $n=N-\ell$. Recall that
\begin{equation}
\begin{aligned}
    \FFint^{(\ell)}_n = \left(\sum_{\text{planar cut}}\mathrm{Cut}_n\right) \mathcal{I}_{N=n+\ell}\ ,\quad \text{where}\quad \mathcal{I}_{N} = \sum_{i}c^{(i)}_{N}\mathfrak{M}^{(i)}_{N}.\\
\end{aligned}
\end{equation}
The above equations indicate two perspectives: (1) cutting propagators of the master diagrams in momentum space, (2) taking light-like limit in the dual coordinate space. We can calculate $\FFint^{(\ell=N-n)}_n$ with $N=3,4,5,6$ and $2\leq n\leq N$ for a given $N$ in both ways. We have put all the result of $\FFint^{(\ell=N-n)}_n$ for all $N=3,4,5,6$ and $2\leq n\leq N$ in the ancillary file. In this subsection, we explicitly show some of them.

\paragraph{Sudakov FF (Two-point cases)} 
Note that
\begin{equation}
    \FFint_{2}^{(\ell)}= \FFint_{2,0}^{(\ell)}=\sum_{\ell_L + \ell_R = \ell} {F}_{2,0}^{(\ell_L)} {\bar{F}}_{2,0}^{(\ell_R)} = \mathrm{Cut}_2\left( \mathcal{I}_{N = 2+\ell} \right),
\end{equation}
from which we can extract the tree$\times$$\ell$-loop part:
\begin{equation}
    \left. \mathrm{Cut}_2\left( \mathcal{I}_{N = 2+\ell} \right)\right|_{\text{tree} \times \ell\text{-loop}} = {F}_{2,0}^{(\ell)} {\bar{F}}_{2,0}^{(0)} + {F}_{2,0}^{(0)} {\bar{F}}_{2,0}^{(\ell)} = 2 \, \FFint_{2}^{(0)} \left( \frac{{F}_{2,0}^{(\ell)}}{F_{2,0}^{(0)}} \right) = \frac{2}{(q^2)^2} {\tilde{F}}_{2,0}^{(\ell)},
\end{equation}
where $\FFint_{2}^{(0)} = 1/(q^2)^2 $ and we have defined the integrand after factoring out the tree-level MHV FF:
\begin{equation}
    {F}_{n,k}^{(\ell)} = F_{n,0}^{(0)} \, {\tilde{F}}_{n,k}^{(\ell)}.
\end{equation}
Thus, the Sudakov FF at $\ell$-loop is
\begin{equation}
\begin{aligned}
    {\tilde{F}}_{2,0}^{(\ell)} & = \frac{1}{2}(q^2)^2 \big{[} \left. \mathrm{Cut}_2\left( \mathcal{I}_{N = 2+\ell} \right) \big{]} \right|_{\text{tree} \times \ell\text{-loop}}  \\
    & = \frac{1}{2}(q^2)^2 \lim_{x_{12}^2,x_{21^{+}}^2 \rightarrow 0} \left. \left[ \left( x_{12}^2x_{21^{+}}^2 \right) \mathcal{I}_{N=2+\ell} \right] \right|_{\text{tree} \times \ell\text{-loop}},
\end{aligned}
\end{equation}
where the first equality is understood to cut $2$ propagators of the master diagrams and the second equality is to take light-like limit in the dual coordinate space.

For $N=3$, we obtain
\begin{equation}
\begin{aligned}
    {\tilde{F}}_{2,0}^{(1)} & = \frac{1}{2} (q^2)^2 \times 2 \times \frac{1}{4} \times  \sum_{\text{cyc}(1,2)} \frac{1}{q^2} \left( \vcenter{\hbox{\scalebox{.6}{\sudakovcutthreeA}}} + \vcenter{\hbox{\scalebox{.6}{\sudakovcutthreeB}}} \right) = \, q^2 \vcenter{\hbox{\scalebox{.7}{\sudakovone}}}\\
    & = \frac{1}{2} q^2 \left( \frac{1}{x_{13}^2 x_{23}^2 x_{1^{+}3}^2} + \frac{1}{x_{23}^2 x_{1^{+}3}^2 x_{2^{+}3}} \right) = \frac{1}{2} q^2 \left( \vcenter{\hbox{\scalebox{.7}{\triA}}} +\vcenter{\hbox{\scalebox{.7}{\triB}}} \right),
\end{aligned}
\end{equation}
where we show the explicit cutting procedure in the first equality. From the above equation, we see that two seemingly different integrands $\vcenter{\hbox{\scalebox{.5}{\triA}}}$ and $\vcenter{\hbox{\scalebox{.5}{\triB}}}$ in the dual coordinate space are actually the same one $\vcenter{\hbox{\scalebox{.5}{\sudakovone}}}$ when written in momentum space. So for compactness, we only show the results in momentum space by drawing the Feynman integrals in the following. For $N=4$, we have
\begin{equation}
\begin{aligned}
    {\tilde{F}}_{2,0}^{(2)}  = & \frac{1}{2} (q^2)^2 \sum_{\text{cyc.}}\left[ 
    2 \times \frac{1}{4} \left(\hspace{-0.5em} \vcenter{\hbox{\scalebox{.6}{\sudakovcutFourA}}} + \vcenter{\hbox{\scalebox{.6}{\sudakovcutFourB}}}\hspace{-0.5em}\right) 
     \right.\\
    & + 2 \left.\times \frac{1}{4} \left(\hspace{-0.5em}\vcenter{\hbox{\scalebox{.6}{\sudakovcutFourC}}} \hspace{-0.5em} + \vcenter{\hbox{\scalebox{.6}{\sudakovcutFourD}}}\hspace{-1em} \right) + 2 \times \frac{1}{8} \left(\hspace{-0.5em}\vcenter{\hbox{\scalebox{.6}{\sudakovcutFourE}}} \hspace{-0.5em} + \vcenter{\hbox{\scalebox{.6}{\sudakovcutFourF}}}\hspace{-0.5em}\right)
    \right]\\
    = & \frac{1}{2} (q^2)^2 \left( 4 \, \vcenter{\hbox{\scalebox{.6}{\sudakovTwoA}}} + \vcenter{\hbox{\scalebox{.6}{\sudakovTwoB}}} \right),
\end{aligned}
\end{equation}
which is the same as Eq.(4.2) in \cite{Boels:2012ew}. And for $N=5$, the cutting procedure is too complicated to write down, so we only show the final result as follow:
\begin{equation}
\begin{aligned}
     {\tilde{F}}_{2,0}^{(3)} & =\frac{1}{2}(q^2)^2 \left[ 8 \, q^2 \vcenter{\hbox{\scalebox{.6}{\sudakovThreeA}}} + 4 \, q^2 \left(\vcenter{\hbox{\scalebox{.6}{\sudakovThreeB}}} +  \vcenter{\hbox{\scalebox{.6}{\sudakovThreeC}}} \right) \right.\\
     & +2 \, \vcenter{\hbox{\scalebox{.6}{\sudakovThreeD}}} \hspace{-0.5em} + 2 \left( (\ell_1 + \ell_2)^2 \, \vcenter{\hbox{\scalebox{.6}{\sudakovThreeE}}}\hspace{-0.5em} + (\ell_1 + \ell_2)^2 \,\vcenter{\hbox{\scalebox{.6}{\sudakovThreeF}}}+ (1 \leftrightarrow 2)\right)\\
    & \left. - \left( \ell_1^2 \, \vcenter{\hbox{\scalebox{.6}{\sudakovThreeE}}} + 2 \ell_1^2 \, \vcenter{\hbox{\scalebox{.6}{\sudakovThreeG}}} + 2 \ell_2^2 \, \vcenter{\hbox{\scalebox{.6}{\sudakovThreeE}}} + (1 \leftrightarrow 2)\right)\right],
\end{aligned}
\end{equation}
where ``$1 \leftrightarrow 2$" in the last parentheses means the exchange of two external legs. The results are consistent with those in \cite{Boels:2012ew} after performing IBP reduction and being expressed by the same family of master integrals.  The results of $N=6$ in terms of dual coordinates can be found in the ancillary file.

\paragraph{Higher points}Beyond Sudakov FF, we only have the integrand for squared FF, rather than the integrand of FF itself. For $N=4$ and $n=3$, the integrand reads
\begin{equation}
\begin{aligned}
    \FFint_3^{(1)} & =\frac{2}{x_{13}^2x_{21^+}^2x_{1a}^2x_{2a}^2x_{1^+a}^2}+\frac{2}{x_{21^+}^2x_{32^+}^2x_{1a}^2x_{2a}^2x_{1^+a}^2}\\
    & +\frac{2}{x_{13}^2 x_{21^+}^2 x_{1a}^2 x_{3a}^2 x_{1^+ a}^2} + \frac{2}{x_{13}^2 x_{32^+}^2 x_{1a}^2 x_{3a}^2 x_{1^+ a}^2} + \frac{2}{x_{32^+}^2x_{1a}^2x_{2a}^2x_{3a}^2 x_{1^+a}^2}+\widetilde{\text{cyc.}}\\
    & = \frac{2}{x_{13}^2} \vcenter{\hbox{\scalebox{.7}{\tridualA}}} + \frac{2}{x_{32^+}^2} \vcenter{\hbox{\scalebox{.7}{\tridualA}}} \\
    & + \frac{2}{x_{21^+}^2} \vcenter{\hbox{\scalebox{.7}{\tridualB}}} + \frac{2}{x_{32^+}^2} \vcenter{\hbox{\scalebox{.7}{\tridualB}}}+ \frac{2}{x_{32^+}^2} \vcenter{\hbox{\scalebox{.7}{\boxdual}}} +\widetilde{\text{cyc.}},
\end{aligned}
\end{equation}
where the first four terms are ``triangles'' and the last term is a ``box'', which we found consistency with Eq.(4.23) in \cite{Bianchi:2018peu}. For higher points, there will be more complicated integrals and products of two lower loop integrals. The calculation in dual space is straightforward, we put all these explicit results in the ancillary file.

\section{Application: Energy Correlators}
\label{sec:EEC}

Form factor matrix elements are key to the study of observables in physical processes at particle colliders or in conformal field theory \cite{Hofman:2008ar}.  Such observables probe the properties of states that are created due to localized excitations in the field theory (sourced by operator $\cO$).  The simplest examples are the $N$-point Energy Correlators ($\rm{E^NC}$), which are IR safe and well-defined in perturbative quantum field theory.

The $\rm{E^NC}$ is formulated in terms of correlation functions of light-transformed stress energy tensor \cite{Kravchuk:2018htv,Kologlu:2019bco,Kologlu:2019mfz},
\begin{align}
{\rm{E^N C}} (z_1 \ldots z_N)  =\frac{ \langle \cO^\dagger \cE (z_1)\cdots \cE (z_N)  \cO   \rangle}{ \langle \cO^\dagger  \cO \rangle }  \, , \quad 
\cE(z) =  \int_{-\infty}^\infty d v\, T_{v v} (0, v , \vec{z} ) .  
\end{align}
The operator $\cE(z)$, acting on free asympotic states, produces the energy flow which goes into the direction of a null vector $z^\mu =(1,|z|^2, \vec{z})$, 
\begin{align}
\cE(z) |p_i \rangle  =  p_i^0\, \delta^2 (z  - \hat{p}_i )  \, |p_i \rangle .
\end{align} 
Hence alternatively, energy correlators are defined
as on-shell phase-space integration over form factor square weighted by a measurement function
\begin{align} \label{F2toENC}
{\rm E^NC} (z_1 \ldots z_N )=  \sum_{n>N} \sum_{\sigma \in S_n} \int {\rm dPS}_{n}  \prod_{i=1}^N \, [p_i^0\, \delta^2 (z_i  - \hat{p}_i ) ] \times  |F_{\cO} (\sigma(1), \ldots, \sigma(n) )|^2  .
\end{align}
In recent years, much progress has been made in calculating two-point energy correlators both in  $\cN=4$ SYM and QCD at higher-loop order.  Relatively little is known for the $\rm{E^NC}$ at generic scattering angles beyond the level of $N=3$ at leading order.
The integrands for the form factor squared will be important ingredients for carrying on the studies on energy correlators towards higher-point/higher-loop level. 
With current data,  the $(5-N)$-loop integrands are available for the $\rm E^{N}C$.  
Towards the integrated functions, investigating the structures of parameter-space integrals in \eqref{F2toENC} is a crucial step.
It would be desirable to exploit the methods developed for Feynman loop integrals, within the framework of Landau/Schubert analysis \cite{Fevola:2023kaw,Fevola:2023fzn,Correia:2025yao,Bourjaily:2020wvq,Britto:2023rig,Dennen:2016mdk,He:2023umf,Yang:2022gko,He:2022tph,He:2022ctv}, projective geometry \cite{Arkani-Hamed:2017ahv,Bourjaily:2019hmc,Gong:2022erh} and intersection theory \cite{Mastrolia:2018uzb,Mizera:2019vvs,Caron-Huot:2021xqj,Caron-Huot:2021iev}.

\paragraph{\texorpdfstring{$N$}{N}-point energy correlator at tree level}

\begin{figure}[htbp]
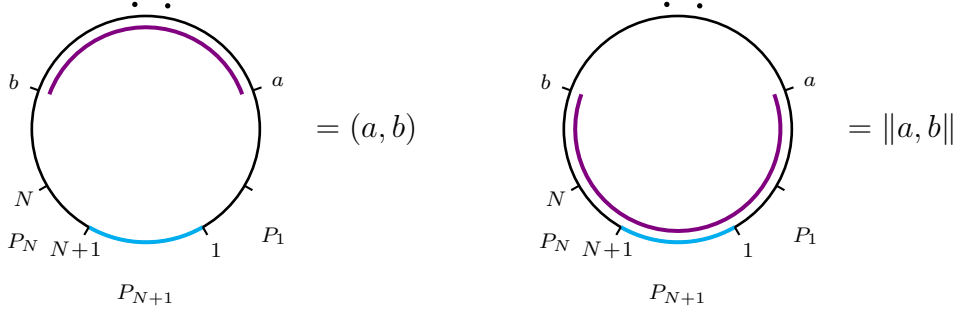

\centering
\mandelstampic
\caption{Mandelstam invariants $ s_{a \cdots (b-1)} = (a, b), \;  s_{b\cdots {\sblue{N+1}} \cdots (a-1)} = \Vert a,b \Vert $  visualized as distance between dual points in a periodic coordinate space with period $q= \sum_{i= 1}^{N+1} p_i $. }
    \label{fig: App1}
\end{figure}

The simplest cases of study are the $\rm{E^N}C$ at LO corresponding to $n=N{+1}$ in \eqref{F2toENC}. 
Their integrands derive from the color-ordered tree-level $\FFint^{(0)}_{N+1}$.  The latter depend on $N$ external on-shell momenta aligned with the direction of energy flow operators $\mathcal{E}(z_i)$, as well as the $(N{+}1)-$th recoiling on-shell momentum.   To specify the kinematics, one may pick a reference frame where 
\begin{align} \label{ENCkinematics}
p_i = \frac{x_i}{1+|z_i|^2}  \begin{pmatrix}
1 & z _i\\
 \bar{z}_i &\, |z_i|^2
\end{pmatrix}  ,  \quad  
 q=  \begin{pmatrix}
1 & \; 0 \\
 0 & \; 1
\end{pmatrix}  , \quad i = 1, \cdots, N
\end{align}
Within this frame let us rewrite 
$\FFint^{(0)}_{N+1}$ in terms of the on-shell degrees of freedom, namely $N$ set of energy and angle variables $(x_i, \vec{z}_i)$. 

For clarity we define two types of inner products. They represent squared distance between dual points in periodic coordinate space, which are visualized on a cylinder in Fig.~\ref{fig: App1}.
\begin{align}
(a, b) := p^2_{a\cdots (b-1)}  , \quad     \Vert a, b \Vert := (b,  \bar a) = (\underline{b}, a),  
  \quad  N+1  \geq b \geq a+1  \geq 2 \, ,
\end{align}
where  $\bar a$ ($ \underline{a}$) is the image of dual point $a$ shifted into the next (previous) period, 
\begin{align}
 \bar a := a+N+1,\;   \underline{a} := a-N-1\,,  \;   \bar a -a =a- \underline{a}   := q \,.    
 \end{align}  
The local Mandelstam invariants in $\FFint_{N+1}$ all fall into the two types, namely $ s_{a \cdots (b-1)} = (a, b), \;  s_{b\cdots {\sblue{N+1}} \cdots (a-1)} = \Vert a,b \Vert $, 
which are parametrized differently 
\begin{align} 
(a, b)  =\frac12 \sum_{i,j \in [a,b-1]}  x_i x_j  \gamma_{ij}\,  , \quad \; \Vert a, b \Vert = 1-  x_{a \cdots (b-1)}  - (a, b) \,. 
\end{align}
where $x_{a\cdots b}$ is a shorthand notation for $\sum_{i \in \{a \cdots b\}} x_i$,
 $\gamma_{ij}$ is the PSL(2) invariant distance between two points $z_i$ and $z_j$ on celestial sphere, 
\begin{align}
\gamma_{ij} := \frac{|z_{ij}|^2}{(1+|z_i|^2) (1+|z_j|^2) }.
\end{align} 
In the meantime, the $(N{+}1)$-body on-shell phase-space measure takes the following form
\begin{align}
{\rm d PS}_{N+1} = \prod_i\, \frac{d^2 z_i}{(1+|z_i|^2)^2} \,[ x_i d x_i ] \,   \delta(\Vert 1, N{+}1 \Vert ) ,
\end{align}
which contains a delta function setting the recoiling $(N{+}1)$-th momentum on-shell.

At LO, angular degrees of freedom in the phase space are fully localized. ${\rm E^N C}$ at LO is an $N$-fold integration over energy, whose integrand is mostly identical to the ratio function for the $(N+1)$-point form factor squared: 
\begin{align} \label{ENCIntegral}
& {\rm E^NC}|_{\rm LO} =\frac{[(1+|z_1|^2)(1+|z_N|^2)]^{-1}}{|z_{12}|^2 \ldots |z_{N-1,N}|^2 } \int_{0}^1 \prod_i [d x_i] \,   I_{\rm E^NC} (x_i ,z_i )   \;\;  + \text{perm} (z_1 \ldots z_N)   ,
\end{align}
where
\begin{align}
  I_{\rm E^NC} (x_i ,z_i )  =   \frac{  x_1 x_N\,   \delta(\Vert 1, N{+}1 \Vert )}{ \Vert 1, N  \Vert  \,  \Vert 2,  N{+}1 \Vert  } \left[ \frac{\FFint^{(0)}_{N+1}}{ \FFint^{(0)}_{N+1,0} }\right].
\end{align}
In particular, on the support of the $\delta$-function, the MHV piece factorizes in terms of two linear propagators, 
\begin{align}
 \frac{ \Vert 1, N  \Vert }{x_N} = 1-   
 \sum_{i=1}^{N-1}  x_i  \gamma_{iN},  \quad 
\frac{ \Vert  2, N{+}1   \Vert }{x_1} = 1-  \sum_{i=2}^{N}  x_i \gamma_{1i}  \nn .
\end{align}

\subsection{Dlog basis for the \texorpdfstring{$\rm{E^3C}$}{E3C} at generic angle}

Singularities in $ I_{\rm E^NC} $ define a set of hyperplanes and quadratic surfaces which intersect with the integration contour
 $\{ \Vert 1, N {+}1 \Vert =0 \, \big|\,  x_i >0, \,  i \in [1,N] \} . $ In particular the ${\rm E^NC}$ integrals are finite. 
It is suitable to apply the method of relative twisted intersection theory \cite{Caron-Huot:2021xqj,Caron-Huot:2021iev}, which allows setting up the calculation without introducing dimensional regularization. As proof-of-principle we will investigate in detail the family of $\rm{E^3C}$ integrals whose integrand derives from $\FFint_4$, where all singularities (except from the $\delta$-function) are hyperplanes. The integrand can be cast into the following form

\begin{align}
 I_{\rm E^3C} &=    \delta(S_{\sblue{4}} )  
  \left[ \frac{1}{T_1 \, T_2 }
+ \frac{S_{\sblue{1}} S_{\sblue{3}} }{T_3 \, T_4 }  +
 \frac{ S_{\sblue{1}}   S^2_{\sblue{2}} S_{\sblue{3} }  \gamma_{12}  \gamma_{23} }{ T_1 \, T_2 \, T_3 \, T_4 } \right.  \nn \\
 & + \frac{ S_{\sblue{1}} S_{\sblue{2} }  S_{\sblue{3}}   \gamma_{12} }{ T_1 \, T_3\, T_4  }  
 + \frac{ S_{\sblue{1}} S_{\sblue{2} }  S_{\sblue{3}}   \gamma_{12} }{ T_1 \, T_3\, T_6  } +   
  \frac{ S_{\sblue{1}} S_{\sblue{2} }  S_{\sblue{3}}   \gamma_{12} }{ T_1 \, T_3\, T_5  } +
 \frac{ S_{\sblue{1}} S_{\sblue{2} }  S_{\sblue{3}}   \gamma_{23} }{ T_2 \, T_3\, T_4  }  
 + \frac{ S_{\sblue{1}} S_{\sblue{2} }  S_{\sblue{3}}   \gamma_{23} }{ T_2 \, T_4\, T_6  } 
 + \frac{ S_{\sblue{1}} S_{\sblue{2} }  S_{\sblue{3}}   \gamma_{23} }{ T_2 \, T_4\, T_5  }     \nn \\
 &\left.  
 + \frac{ S_{\sblue{1}} S_{\sblue{2} }  S_{\sblue{3}}   \gamma_{12}}{ T_1 \, T_5 \, T_6 }    +
   \frac{ S_{\sblue{1}} S_{\sblue{2} }  S_{\sblue{3}}   \gamma_{23}  }{ T_2 \, T_5 \, T_6 }   
  + \frac{  S_{\sblue{2}}  S^2_{\sblue{3}}   \gamma_{23}   }{T_1 \, T_5 \, T_6 }  + 
   \frac{ S^2_{\sblue{1}}  S_{\sblue{2}}   \gamma_{12} }{T_2 \, T_5 \, T_6 }  
  \right].
\end{align}
Singularities that appear are grouped into two sets $\vec{T} = \{T_1, \cdots, T_6 \}$  and $\vec{S} = \{S_{\sblue{1}}, \cdots S_{\sblue{4}} \} $,
\begin{align}
 T_{1}= 1-x_2 \gamma_{12} - x_3 \gamma_{13} \, , \quad 
  T_{2} =  1-x_1 \gamma_{13} - x_2 \gamma_{23} \, ,   \nn \\
T_3 = 1-x_1\, , \quad  T_4 =1-x_3\, ,  \quad T_5 =1-x_1-x_2-x_3\,,  \quad T_6 = 1- x_2\,,  \nn \\
S_{\sblue{1,2,3}} = x_{1,2,3},  \quad  S_{\sblue{4}} =  1- x_1- x_2 -x_3 + x_1 x_2 \gamma_{12}+ x_{1} x_3 \gamma_{13} +x_2 x_3 \gamma_{23}  ,
\end{align}
where $S_{\sblue{1}, \sblue{2}, \sblue{3}} $ are the boundaries that define the integration contour. They do not appear explicitly on the denominators of the $\rm{E^3C}$ integrand thanks to the finiteness property of the observable. 

The singularity boundaries $\vec{T}$ and $\vec{S}$  define a cohomology.  
Following the approach in \cite{De:2023xue}, 
one may pick a subset of $\vec{T} \cup \vec{S}$ and 
set them as relative twisted boundaries. 
In our case since the integrand is free from end-point singularities at $\{ x_i =0\}$, we can make any desirable choice of twisted versus untwisted boundaries.  In the following, we will interpret $\vec{T}$ as twisted boundaries and $\vec{S}$ as relative boundaries and construct the basis of a dual cohomology.   
In particular, we will exert boundary stratification on $\vec{S}$ and 
build a basis of dual forms that live on the locus  $\bigcap_{\sblue{i} \in \sblue{J}}  \{ S_{\sblue i} =0 \}, \forall\, {\sblue{J}}  \supseteq   \{ {\sblue{4}} \}$. Altogether they form a complete dlog basis for the ${\rm E^3C}$, whose transcendental weights are graded by the codimension of boundary surfaces they live on.

\begin{figure}[htbp]
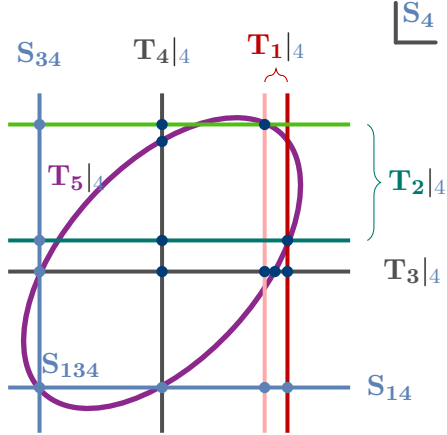

    \centering
    \singulatiryarrangement
    \caption{  Arrangement of singularity hyperplanes $T_{1- 5}$ intersecting on the $S_{\sblue 4}-$boundary projected onto the $(x_1, x_3)$ plane, where $T_1|_{\sblue{4}}$ ($T_2|_{\sblue{4}}$) appears as a pair of straight parallel lines.  The number of intersecting points between pairs of singularity curves $  \left\{ (T_1 |_{\sblue{4}}, T_2 |_{\sblue{4}}), \,  (T_1 |_{\sblue{4}} , T_3  |_{\sblue{4}}),\, (T_3  |_{\sblue{4}}, T_4  |_{\sblue{4}} ),  \, (T_1 |_{\sblue{4}}, T_5  |_{\sblue{4}}),\,  (T_3 |_{\sblue{4}}, T_5 |_{\sblue{4}}) \right\}   $ are given by $\{ 2, 2, 1, 2, 2 \} $, respectively. 
     }
\end{figure}

Let us start with the codimension-1 relative boundary $S_{\sblue{4}}$. Among the twisted boundaries,  we identify 5 pairs of hyperplanes:  
\begin{align}
 \left\{ (T_1 , T_2 ), \,  (T_1  , T_3  ),\, (T_3  , T_4  ),  \, (T_1 , T_5  ),\,  (T_3 , T_5 ) \right\}   \nn 
 \end{align}
   which intersect on $S_{\sblue{4}}$, thus generating 5 top topologies for the $\rm{E^3C}$ (up to dihedral flip). 
Within these topologies, we construct a basis of 11 independent dlog forms on $S_{\sblue{4}}$, which evaluate to polylogarithms with transcendental weight two.  
These forms are given explicitly as follows: 
\begin{align} \label{dlog11}
\varphi_1 &= \delta_{\sblue{4}}  \left[  {\rm d} \log T_1 |_{\sblue{4}} \wedge {\rm d} \log  T_2|_{\sblue{4}}  \right]  ,  \quad 
\varphi_2 = \delta_{\sblue{4}}  \left[  {\rm d} \log T_1 |_{\sblue{4}} \wedge {\rm d} \log  T_3|_{\sblue{4}}  \right]  ,  \nn \\
\varphi_3 & = \delta_{\sblue{4}}  \left[  {\rm d} \log T_3 |_{\sblue{4}} \wedge {\rm d} \log  T_4|_{\sblue{4}}  \right]  ,\quad  
\varphi_4 = \delta_{\sblue{4}}  \left[  {\rm d} \log T_1 |_{\sblue{4}} \wedge {\rm d} \log  T_5|_{\sblue{4}}  \right]  , \nn \\
\varphi_5 & = \delta_{\sblue{4}}  \left[  {\rm d} \log T_3 |_{\sblue{4}} \wedge {\rm d} \log  T_5|_{\sblue{4}}  \right]  ,\quad  
\varphi_6  = \delta_{\sblue{4}}  \left[ \frac{ \gamma_{13} \Delta_1 {\rm d}^2 x }{T_1 T_2} \right]  ,  \nn \\
\varphi_7  & = \delta_{\sblue{4}}  \left[ \frac{\Delta_1 {\rm d}^2 x }{T_1 T_3} \right]  , \quad 
 \varphi_8 = \delta_{\sblue{4}}  \left[ \frac{ \gamma_{23} {\rm d}^2 x }{T_3} \right]   ,  \quad  
 \quad  \varphi_9   = \delta_{\sblue{4}}  \left[ \frac{ \Delta_1 {\rm d}^2 x }{T_1 T_5} \right] \,, \nn\\
\varphi_{10}  &= \delta_{\sblue{4}}  \left[ \frac{ (\gamma_{12}-\gamma_{13}) {\rm d}^2 x }{T_3 T_5} \right]  , \quad \varphi_{11}  = \delta_{\sblue{4}}  \left[ \frac{ \Delta_2 {\rm d}^2 x }{ T_5} \right]  
\end{align}
where $\delta_{\sblue{i}}$ indicates that the forms are defined on the vanishing locus of $S_{\sblue{i}}$, more explicitly, 
\begin{align} 
\delta_{\sblue{i}} [ \cdot ] := {\rm Res}_{S_{\sblue{i}}=0} \left[ {\rm d} \log S_{\sblue{i}} \wedge \cdot \,  \right]  \,,   \quad  \delta_{\sblue{i|j}} :=   \delta_{\sblue{j}} \delta_{\sblue{i}}   
\end{align}
The leading singularities of the basis contain two types of square roots: 
\begin{align}
\Delta_1=\sqrt{   4 \gamma_{12}\gamma_{13}\gamma_{23} + \Delta_2^2},  \quad   \Delta_2 = \sqrt{ \gamma_{12}^2 +  \gamma_{13}^2 + \gamma_{23}^2 - 2 \gamma_{12} \gamma_{13}-2 \gamma_{12} \gamma_{23} - 2\gamma_{13} \gamma_{23} } \nn \, .
\end{align}
We verify that  $\{\varphi_1, \cdots \varphi_{11}\}$  can be mapped onto the basis of weight-two function $\{g_1, \cdots g_{11}\}$ introduced in \cite{Yan:2022cye}. They couple, through differential equations, to forms that live on lower dimensional boundaries. Next,  we move onto a codimension-$2$ boundary $S_{\sblue{4}} \cap S_{\sblue{1}}$, where one can write down 3 independent one-forms : 
\begin{align}\label{dlog3}
\varphi_{12} = \delta_{\sblue{4|1}} \left[  {\rm d} \log T_1|_{\sblue{41}} \right]  , \quad 
 \varphi_{13} = \delta_{\sblue{4|1}} \left[  {\rm d} \log T_4 |_{\sblue{41}} \right]  , \quad 
  \varphi_{14} = \delta_{\sblue{4|1}} \left[  \frac{\Delta_1 {\rm d} x}{T_1} \right]
\end{align}
which agree with the weight-one function $\{f_1 ,f_2, f_3 \}$ in \cite{Yan:2022cye}. Finally on the codimension-$3$ boundaries $S_{\sblue{4}} \cap S_{\sblue{i}} \cap S_{\sblue{j}}$, there exist zero-forms, \textit{e.g.} 
\begin{align}
\varphi_{15} = \delta_{\sblue{4|1|3}} [1]
\end{align}
which appear in the system of differential equations, although weight-zero terms are absent from the final result for ${\rm E^3C}$. 
The symbol alphabets for the dlog forms in \eqref{dlog11} and \eqref{dlog3} can be computed through direct integration via {\textsc{HyperInt}}\cite{Panzer:2014caa}  or differential equations \cite{Gong:E4C}.
In general, the results obtained for the $\rm{E^N}C$ alphabets will offer valuable reference for developing the framework of Landau analysis for physical observables.

\subsection{A classification of the elliptic \texorpdfstring{$\rm{E^NC-}$}{ENC}integrals }

Singularities that appear in the ${\rm E^NC}$ are summarized in the following, where $T_{\blue{ab}}, T_{\red{AB}}, S_{\sblue{N+1}} $ are quadratic while the remaining ones are linear.  The kinematic dependence is contained in the homogeneous function $(a,b) = \frac12 \sum_{i,j \in [a,b{-}1]}  x_i x_j \gamma_{ij} $. 
\begin{align} \label{ENCsingularity}
&T_1  = \frac{\Vert 1, N \Vert }{x_1} \,, \qquad  T_2 =  \frac{\Vert 2, N{+}1 \Vert }{x_N}  \,, \quad   \\
& T_{\blue N+1}  = (1, N{+}1 ) = 1-  x_{1 \cdots N}\,,    \quad    T_{\red A}   = ||a , a+1 || =   1- x_a , \quad   a \in [1, N] \,,       \nn \\
&T_{{\blue ab}}  = (a, b) \,,            \qquad  
3 \leq b-a \leq N{-}1 \,, \;  a,b  \in [1,  N{+}1] \,,  \nn \\
 & T_{{\red AB}}  = \Vert a, b  \Vert  =  1- x_{a \ldots (b{-}1)} +(a, b)\,,  \quad   
 \quad  
  2 \leq b-a \leq N{-}2\,,  \; a,b  \in [1,  N{+}1] \,, \nn \\
 & S_{\sblue{1,2, \cdots,N}} = x_{1,2, \cdots,N} \,, \quad  S_{\sblue{N+1}}  = \Vert 1, N{+}1 \Vert =  1-  x_{1 \cdots N} +  (1,N{+}1) \, . \nn 
\end{align}

The analysis previously carried out for the ${\rm E^3C}$ could in principle be generalized to higher point energy correlators with arbitrary $N$.  However, starting from $N=4$, we observe two irreducible quadratic singularity surfaces intersecting each other on the vanishing locus of $ S_{\sblue{N+1}} $. (A similar situation takes place in the multi-collinear limit of $\rm{E^NC}$ starting from $N=5$ \cite{He:2024hbb}).  
In consequence  ${\rm E^NC}$ is no longer purely polylogarithmic.  The space of the functions for the four-point energy correlator at generic angle is given in \cite{Gong:E4C}. The analysis is carried out by means of Picard-Fuchs differential equations~\cite{Muller-Stach:2012tgj,Adams:2017tga,Dlapa:2020cwj,Dlapa:2022wdu}, whose structure suggests 3 types of elliptic curves of genus 1 and 2 \cite{Bourjaily:2018ycu}.

Given the increasing number of kinematic variables,  and the fact that elliptic curves emerge at level of $N=4$,  we expect the analytic structures for the ${\rm E^NC}$ to be rather intricate. Considering the scope of this paper, we will only give a set of {\it{necessary}} conditions for an integral topology to be {\it non-polylogarithmic}, for arbitrary $N$. We will postpone more systematic analysis on the function space to future work.

For simplicity, we will focus on integral topologies featuring double cuts $\{T_I , T_{II} \}$.  (We will comment on integrals with more cuts later on.)  $ \{T_I , T_{II} \}$  are drawn from the singularity boundaries in \eqref{ENCsingularity} intersecting on $S_{\sblue{N{+}1}}$.  
In order for the topologies to be non-polylogarithmic, they need to satisfy a set of criteria which we summarize in the following. 

\begin{itemize} 
\item[{\it 1}]. $\{ T_{I} , T_{II} \}$ must depend on all $N$ integration variables $\{ x_1,  \cdots, x_N \} $.  \label{statement1}

If this is not the case, then let us assume $x_a$  appears only in  $S_{\sblue{N{+}1}}$, not in $T_I$ or $T_{II}$.  If  $T_I, T_{II}$ are both quadratic and irreducible, then we can always find another variable $x_b$ which appears in $T_I$ but not $T_{II}$. Hence equations $T_{I} = S_{\sblue{N{+}1}} =0$  can be solved linearly in terms of $(x_a, x_b)$ without increasing the polynomial degree of $T_{II}$. 
The solution defines a hypersurface which is degree-2 at most, namely not elliptic. 

\item[{\it 2}]. If  $T_{I}= (a,b)$, which is homogeneous in $x$, then $T_{II}$ must overlap with $T_I$, i.e.  $T_{I}$ and $T_{II}$ share at least one variable $x_a$ .  In particular, 

\subitem{\it a}. if $T_{II}$ is also homogeneous, then it must only partially overlap with $T_I$.
  
\subitem{\it b}. if $T_{II}= \Vert c, d \Vert$, which is not homogeneous, then it must share at least two variables with $T_{I}$.

If $T_{I}$ and $T_{II}$ do not overlap, then let us assume $T_{I}$  depends on  $X= (x_i, \cdots, x_j \cdots ) $, while $T_{II}$ depends on $Y = (y_i , \cdots)$.  Once we change variables into $X  \rightarrow  x_i (1, \cdots, \bar{x}_j, \cdots ) $, the $x_i-$dependence in $T_I$ factors out.  In the meantime $x_i$ appears linearly in $S_{\sblue{N}{+}1} |_{T_I =0}$.  Now we have the exact same situation as previously discussed in ({\it 1}) and therefore statement  ({\it 2}) is true.   Statement  ({\it 2a}) follows from a similar argument:  in the case where $Y \subset X $, we can change variables into $X \rightarrow  x_i (1, \cdots, \bar{y}_i, \cdots ) , Y \rightarrow x_i  (\bar{y}_i , \cdots)$ .  

Next we consider statement  ({\it 2b}). Let us assume $X \cap Y = \{x_i\}$ and $ X \cup Y = \{x_1, \cdots, x_N \}$.   Apparently, $S_{\sblue{N{+}1}} - T_I -T_{II}$ does not depend on $x_i$. Again we can apply statement (1) by change of variables into $X \rightarrow  x_i (x_a, \cdots, 1, \cdots )$ and argue that $x_i$ only appears in one cut equation $T_{II}=0$.

\item[{\it 3}].  $\{ T_{I} , T_{II} \}$  cannot be drawn from $T_{1}, T_2 $ or $T_{\red A}$. 

For instance consider a counterexample where $T_{I} = T_1 \equiv  -1+ \sum_{i=2}^N  \gamma_{1i}\,  x_i$.  In this case neither $T_I$ nor $S_{\sblue{N}{+}1} |_{T_I =0}$  depend on $x_1$.  We are again in the similar situation as statement  ({\it 1}). 

\end{itemize}

Based on the statements above,  $\{ T_{I}, T_{II} \}$ that give rise to  (beyond) elliptic functions must correspond to pairs of incompatible cuts (in terms of the Mandelstam $s_{ij}$'s). Moreover,  such pairings can be divided into three categories,  each characterized by two segments of dual points lying on one period($= N{+}1$ points) in dual space. 

\paragraph{Type I: }  $\left\{ (1, a) , \; (b, N{+}1) \right\} , \quad b\leq a-1 .$

\paragraph{Type II: }  $\{ (1, a) , \; \Vert b, N{+}1 \Vert \} $ ~(or  $\{   (b, N{+}1 ),   \;   \Vert 1, a \Vert \} $),  ~~~$ b\leq a-2,$ ~
 including the special case where $a=N+1$ (or  $b=1$).

\paragraph{Type III: }  $\{ \Vert 1, a \Vert , \; \Vert b, N{+}1 \Vert \} , \quad b \leq a .$

\vspace{6pt}

Integrals that contain such cut configurations are potentially elliptic, since the cut equations ${ T_I = T_{II} =S_{\sblue{N{+}}1}}$ define a higher degree ($=4$) surface.  

For our purpose we will look into the maximally and minimally overlapping configurations within each type, for the first time it appears in the series of ${\rm E^NC}$( for arbitrary $N$). More specifically, we consider the following cases: 
\begin{itemize}
\item Minimally overlapping:  $ a - b \geq l_{\text{min.}} $. 

For {\bf Type I,II,III}, $l_{\text{min.}} = 1, 2, 0$, respectively . 

\item Maximally overlapping:  $  b= b_{\text{min.}} , a = a_{\text{max.}} , \,  a-b >  l_{\text{min.}}$  

For {\bf Type I,II,III}, $(b_{\text{min.}} , a_{\text{max.}} )= (2, N),\, (3 ,N+1), \, (3, N-1)$, respectively. 

\end{itemize}
Integrals in more general configurations typically involve more complicated structures. 
For arbitrary $N$, consider a topology characterized by two segments $[1, a]$ and $[b,N{+}1]$ in dual space.  It can be reduced to an $(N{-}1)$-point configuration once we identify two points $i$ and $j$ lying on either the overlapping segment  $[a,b]$ or non-overlapping segments $[1,a{-}1], [b{+1},N{+}1]$.  The operation is equivalent to taking kinematic limit via sending $\gamma_{ij} \rightarrow 0$. 
This procedure can be repeated iteratively until we reach a set of simplest topologies which appear in the case of $N=4$ or $5$. 
We will go through each of them in the following. 

\vspace{6pt}

For {\bf Type I} integrals, let us make the cut equations projective by introducing a variable $x_0$ through $x_i \rightarrow x_i/x_0$. Taking residues in $x_1, x_N, x_0$, we are left with a square-root leading singularity
\begin{align} \label{eqn:RP6}
\frac{1}{ \sqrt{P_6( x_2, \ldots, x_{b}, \ldots, x_{a{-}1} , \ldots x_{N-1})}}
\end{align} 
where $P_6$ has homogeneous degree 6.  It has maximum degree 6 in terms of the overlapping variables  $x_{b, \cdots, a{-1} } $ and maximum degree 4 in terms of the remaining variables.  In the maximally overlapping configurations, $b=2, a=N$ and $P_6$ is uniformly degree 6 in all projective variables.   Such configuration appears for the first time in the case of  $N=4$, namely $\{s_{123},s_{234}\}$.  Setting $x_2=1$,  $P_6$ is degree 6 in $x_3$, which indicates a hyper elliptic curve of genus 2. The minimally overlapping configuration corresponds to $b=a-1$, which appears for the first time when $N=5$, namely $\{ s_{123}, s_{345} \}$. Setting $x_3=1$, $P_6$ is degree 4 in both $x_2$ and $x_4$, and we cannot take further residues. Nevertheless in the multi-collinear limit, $P_6$ reduces to a perfect square and we can further take residue until we find a standard square root $1/\sqrt{P_4 (x_4)}$, which indicates an eMPL \cite{He:2024hbb}.  

For {\bf Type II}  integrals,   following the same procedure we take triple residues and obtain two types of leading singularities from the double cuts $\{ s_{1 \cdots (a-1)}, s_{{\sblue{N{+}1}} \cdots (b-1)} \} $,
\begin{align}\label{eqn:RQ6}
\frac{1}{\sqrt{Q_6 (x_2, \cdots, x_{N-1})}} ,  \quad  \text{for} \; a  \leq  N\,,  \quad   
 \frac{1}{\sqrt{Q_4 (x_2, \cdots, x_{N-1})}} ,  \quad  \text{for} \; a =  N{+}1
\end{align} 
where $Q_{6(4)}$ is a degree 6(4) homogeneous polynomial.  $Q_6$ and $Q_4$ both have maximum degree 4 in terms of all the remaining $x-$variables.   
For these integrals, minimally overlapping configuration corresponds to $b= a-2$ and maximally overlapping indicates $a= N+1, b= N-1 $. 
Hence both configurations coincide into $\{s_{1234}, s_{{\sblue{5}} 12} \}$  in the case of  $N=4$.  Setting $x_2=1$ in $Q_4$, we obtain a quartic in $x_3$ indicating an eMPL.

For {\bf Type III}  integrals, one may take triple residues in $x_1, x_N, x_{N-1}$ and again find two types of leading singularities 
for the doublets  $\{ s_{a \cdots {\sblue{N{+}1} } }  ,  s_{{\sblue{N{+}1}}  \cdots (b-1)} \} $, 
\begin{align} \label{eqn:RR6}
\frac{1}{\sqrt{R_6 (x_0, x_2 \cdots, x_{N-2})}} ,  \quad  \text{for} \; b  \leq  a-1 \,,  \quad   
 \frac{1}{\sqrt{R_4 (x_0, x_2 \cdots, x_{N-2})}} ,  \quad  \text{for} \; b =  a
\end{align}  
Again, $R_{6(4)}$ is a degree 6(4) homogeneous polynomial.  $R_6$ is degree 6 in terms of $x_0$ and degree 4 in terms of the other variables.

In Type III,  $b=a$ is the minimally overlapping configuration, which appears for the first time in $\{ s_{34{\sblue{5}}}, s_{{\sblue{5}}12} \}$ when $N=4$, whose leading singularity is again a quartic $1/\sqrt{R_4(1, x_2)}$.  
The maximally overlapping configuration, i.e. $b=3, a=N-1, b-a \geq 1 $,  appears for the first time in the case of $N=5$ as $\{ s_{45{\sblue{6}}} ,  s_{{\sblue{6}}12}  \}$ . 
In its leading singularity $1/\sqrt{R_6 (1,x_2, x_3)}$ we could not further take residues.  On the other hand we could test whether this integral topology is polylogarithmic by including one more singularity surface $ \Vert 2 3 \Vert = s_{345{\sblue{6}}1} =1-x_2 $ into the doublet.   Cutting the triplet $\{ s_{45{\sblue{6}}} ,  s_{{\sblue{6}}12} , s_{345{\sblue{6}}1}  \}$  we find a genus-1 elliptic curve.  By virtue of statement  ({\it 3}), the topology  $\{ s_{45{\sblue{6}}} , s_{{\sblue{6}}12}  \}$  cannot be polylogarithmic.

Finally, let us complete the argument by discussing topologies that involve more than two cuts. 
This amounts to adding a set of cuts compatible with $T_I$ or $T_{II}$ to the doublets. 
Looking into the multi-cut equations, we always end up cutting two quadratic terms producing square-root residues in the similar form as Eq.~\eqref{eqn:RP6},\eqref{eqn:RQ6},\eqref{eqn:RR6}. Without aiming for a general proof, we will write down all generalized $M$-cut topologies within {\bf Type I},{\bf II} and {\bf III}, which we conjecture to be elliptic.  
In particular, here we only list cut configurations which are {\it unremovable}: 
 removing any propagator from the set of cuts makes the integral polylogarithmic. 
 They are given as $M$ interlacing segments in dual space :
\paragraph{Type I: }  $\left\{ (1, a_1) , \, (b_2, a_2) , \cdots,  (b_M, N{+}1) \right\}  $
\paragraph{Type II: }  $\left\{ (1, a_1) , \, (b_2, a_2) , \cdots,  \Vert b_M, N{+}1 \Vert  \right\} $  ~ (or $\left\{ \Vert 1, a_1 \Vert, \, (b_2, a_2) , \cdots,  ( b_M, N{+}1 ) \right\} $)
\paragraph{Type III: }  $\left\{ \Vert 1, a_1 \Vert , \, (b_2, a_2) , \cdots,  \Vert b_M, N{+}1 \Vert  \right\} $

\noindent where $\{a_i\}, \{b_i \}$ are length$-M$ increasing sequence satisfying
\begin{align} 
 b_{i-1} < b_i < a_{i-1},  \quad b_{i+1} < a_i < a_{i+1}, \quad a_{i-1} \leq b_{i+1}   ,  \quad b_1 =1, \quad   a_M= N+1\,.  \nn
\end{align}
In addition, when more than 2 cuts are allowed, statement ({\it 1}) needs to be modified and it is indeed possible to have an elliptic cut configuration that involves only $N{-}1$ integration variables. This observation leads to two new types of  $M$-cut topologies, which can be derived from {\bf Type I} or {\bf II}  by setting  $b_1 =2$ (or $a_M = N$), $ \forall \, M \geq 3$.  For example, the following triplet (belonging to Type II) appears both in the case of $N=5$ and $N=6$:
\begin{align}
\{   \Vert 1,3 \Vert,  \, (2,5), \,  (3,6)  \}  = \{ s_{34\cdots {\sblue{N}}}, \, s_{234},\,  s_{345} \}
\end{align}
In the case of $N=5$, the cut equations suggest an eMPL, whereas in the case of $N=6$, it indicates an elliptic curve of genus 2 \cite{Bourjaily:2018ycu,Broedel:2017kkb, Broedel:2019hyg}.  

\vspace{6pt}

In summary, $N$-point energy correlators define a class of $N$-fold iterated integrals depending on  $2N{-}1$ kinematic variables.  
We have classified the {\it non-polylogathmic} topologies for the $N$-point EC into three types. We 
 investigated the simplest 2-cut integral topologies within {\bf Type I, II} and {\bf III},   all of which turn out to be elliptic. 
 More general 2-cut singularity configurations are related to these simple ones by taking kinematic limits. 
 It would be interesting to generalize our analysis to generic $M$-cut topologies. 
We expect that analysis on the elliptic ${\rm E^NC}$-integrals can be done further by applying systematic approaches \cite{Bourjaily:2021vyj,Frellesvig:2021hkr,Broedel:2017kkb,Broedel:2018qkq,Broedel:2019hyg,Bourjaily:2020hjv}. 

\section{Conclusions and Outlook}

In this paper, we have bootstrapped the tree-level form factor squared, $\FFint_N^{(0)}$, in planar ${\cal N}=4$ SYM up to $N=6$, which in fact unifies $\FFint_n$ at $N{-}n$ loops for $2\leq n \leq N$ in a single rational function represented by a collection of two-point master diagrams. As a proof of concept, we have demonstrated how simple the ansatz turns out to be, {\it e.g.} only $4$ and $13$ topologies for $N=5, 6$ cases with $11$ and $163$ parameters, respectively, if we carefully consider power-counting (including ``no triangle") and the ``rung rule"; these parameters can then be easily fixed using various physical limits such as the soft limit, and the multi-collinear limit that reduces $\FFint_N$ to the well-known splitting function. This represents a first step towards more systematic bootstrap program of FF squared at tree and loop level, which, as we have discussed, can provide valuable data with a wide range of applications including energy correlators. 

Our preliminary studies have opened up several new avenues for future investigations, and here we list some of them.  
\paragraph{Graphical bootstrap} An obvious question is how much further we can push the bootstrap for $\FFint_N$, which can be compared with that for amplitude squared via $f$-graphs. The setup here is slightly more complicated but one can generate the analog of $f$-graphs directly in dual space to reasonably high points. For example, by considering embeddings in the dual space, we find that there are $1,3,6,21,72,319,1482,7678,41512$ top topologies for two-point master diagrams compatible with power counting (embedded in the dual space) for $N=3,4, \cdots, 11$; this can be compared with top topologies of vacuum diagrams for amplitude squared, or the ``denominator" graphs of $f$-graphs (which are triangular planar graphs with valency at least $4$): at $6,\cdots, 16$ points (two to twelve loops), there are $1,1,2,5,12,34,130,525,2472,12400,65619$ such top topologies there. Although such counting will significantly increase after dressing with ``numerators", we expect that the ansatz is still manageable on a laptop at least up to $N=9$ or so (comparable with $f$-graphs at nine or ten loops), However, the main obstacle is to derive graphical rules for these graphs, similar to the recently derived ``double-triangle rule" for $f$-graphs~\cite{He:2024cej, Bourjaily:2025iad}, which incorporates and go beyond all the soft limits. This and previous graphical rules~\cite{Bourjaily:2016evz} have allowed one to bypass the computationally challenging issue of representing and constraining rational functions associated with $f$-graphs, enabling the much more efficient {\it graphical bootstrap} for amplitude squared all the way to twelve loops. If the new graphical rule here turns out to be powerful enough just as the one for $f$-graphs, we are confident that such a graphical bootstrap for $\FFint_N$, {\it i.e.} one that does not involve explicit rational functions or conditions such as multi-collinear/soft limits, can succeeded at least through $N=9$ without significant computational resources (which among other things, would contain the integrand of Sudakov FF in the planar limit up to seven loops). 

\paragraph{Geometries underlying (the square of) form factors} It is trivial to extract two-point (Sudakov) FF from the squared object, but it would be very interesting to see how to extract super-FF at higher points, similar to that for amplitudes~\cite{Heslop:2018zut}. In principle we expect that all the information of (loop integrands of) super-FF for any $n,k$ are contained in the corresponding squared objects, although it requires more works for actually extracting higher-$n$ results. The simplest non-trivial case is the three-point FF which are similar to five-point amplitudes (with only MHV and its parity-conjugate $\overline{\rm MHV}$)~\cite{Ambrosio:2013pba, Bourjaily:2016evz}, and it would be interesting to obtain new results for higher-loop three-point FF from the squared objects. One important motivation is to find possible underlying geometries in analogy with the amplituhedron and its square~\cite{Arkani-Hamed:2013jha, Arkani-Hamed:2017vfh, Dian:2021idl, He:2024hbb}: it is tempting to wonder if we could derive these geometries for the FF-squared and even FF themselves, which is already an interesting open question at tree level (see~\cite{Bork:2014eqa} for earlier works on NMHV case). Given earlier works on Grassmannian and loop integrands for FF, these geometries must also allow us to derive all-loop recursion relations for FF (in momentum twistor space), which was considered at one loop in~\cite{Bianchi:2018peu}. Note that given the simplicity of FF squared, it is plausible that one may first find the squared-hedron from our results (similar to the amplituhedron-squared) and then consider refining to certain ``FFhedron", which may shed light on some unified geometric picture underlying various physical quantities, including (square of) amplitudes and form factors, energy correlators and half-BPS correlators, in ${\cal N}=4$ SYM. Given the recent discovery of all-loop amplituhedra in ABJM theory~\cite{He:2022cup, He:2023rou}, it would be interesting to study squared amplituhedron and even certain ``FFhedron" in this theory as well.  

\paragraph{Relations to other quantities, integrations and periods} Recall that for amplitude squared, there are two interpretations for the collection of $N$-point $f$-graphs (dual to vacuum diagrams at $N{-}1$ loops): it serves both as the master diagrams which unify $n$-point amplitude squared at $N{-}n$ loops in various light-like limits (for $4\leq n\leq N$), and also as the $(N{-}4)$-loop integrand of four-point correlator. It would be highly desirable to pinpoint the latter interpretation for the combination of two-point master diagrams,  ${\cal I}_N$, which can not only provide further physical constraints for our bootstrap but also open up more applications and reveal new connections. Moreover, just as the squared amplitudes are dual to Wilson loops in adjoint representation, we expect that any new understanding along this direction will shed new lights on a potential duality between FF-squared and some version of periodic Wilson loops. 

It would be extremely interesting to perform loop integrations for FF-squared, or directly bootstrap the final integrated results. Our current and future data at integrand level provide another playground for the study of planar and non-planar Feynman integrals with nice properties. For example, those integrals for Sudakov FF compute the cusp anomalous dimension in dimensional regularization, and recent works have also unveiled the connection to the octagon~\cite{Coronado:2018cxj,Coronado:2018ypq} by putting the integrands in the Coulomb-branch~\cite{Belitsky:2022itf,Belitsky:2023ssv,Belitsky:2024agy,Belitsky:2024dcf}. Going to higher points, clearly an important question is to see if one can connect these integrands to the remarkable bootstrap program for integrated results for the remainder function at least for $n=3,4$~\cite{Dixon:2024yvq,Dixon:2022rse,Dixon:2020bbt}. In fact, there is even a simpler setting, which is the analog of the ``integrated correlator"~\cite{Binder:2019jwn,Chester:2019pvm,Chester:2020vyz,Dorigoni:2021guq,Dorigoni:2022zcr,Paul:2022piq,Wen:2022oky,Brown:2023zbr}, namely perform loop integrations for all $N$ points which is given by the sum of {\it periods} associated with these two-point diagrams. A computation based on Feynman diagrams have up to three loops has produced remarkably simple results and revealed interesting patterns~\cite{Bianchi:2023llc} and it would be highly desirable to compare with periods of these diagrams (similar to those studied for $f$-graphs {\it c.f.}~\cite{Wen:2022oky,Brown:2023zbr}), which can be pushed to even higher loops. 

\paragraph{Energy correlators and more}
The EC integrands are finite, having a $S_N$-symmetry, behaving nicely in various OPE limits \cite{Kologlu:2019bco,Kologlu:2019mfz,Chang:2020qpj,Chen:2025rjc}. On the other hand, due to partial fraction identities that relate different master diagrams, individual diagram may obscure these physical properties. In this regard the current formalism for the EC integrands is arguably the ideal or minimal one. It would be interesting to search into alternative formalisms for bootstrapping the EC itself without referring to master diagrams.  Progress in this direction may offer insights into the EC in a general context of particle physics and quantum field theory, including QCD, quantum gravity/supergravity \cite{Herrmann:2024yai} and ABJM theory. 
It would also be important to apply the bootstrap approach to event shapes for general detectors where on-shell physical states in the form factor are not inclusively summed over \cite{Korchemsky:2021okt,Belitsky:2013bja,Belitsky:2013xxa}.  
Moreover, as an iterated $N$-fold integral, $N$-point EC is tantalizingly similar to certain loop integrals in Feynman parameter space. 
It would be intriguing to establish a general mapping between EC and Feynman loop integrals (simple cases have been worked out \cite{Chen:2019bpb}) .  
This would ultimately leads to the understanding on the {\it symbology} of the EC, as achieved in the context of Feynman integrals \cite{Goncharov:2010jf,Golden:2013xva,Drummond:2014ffa,Dennen:2015bet,Dennen:2016mdk}.   
Finally, it is prospective to proceed with integrated EC at higher-loop level. Current results on the integrands suggest strongly the development of automated programs for integration by parts in terms of on-shell parameters.

\begin{CJK*}{UTF8}{}
\CJKfamily{gbsn}
\acknowledgments
It is our pleasure to thank Jianyu Gong, Xuhang Jiang, Andrzej Pokraka for inspiring discussions. The work of S.H. has been supported by the National Natural Science Foundation of China under Grant No. 12225510, 12447101, and by the New Cornerstone Science Foundation through the XPLORER PRIZE. K.Y. is supported by the National Natural Science Foundation of China under Grant No. 12357077. 
\end{CJK*}

\appendix

\section{Review on FF squared from BCFW recursion}\label{sec:bcfw}

In this appendix, we review the calculation of tree-level super form factor by BCFW recursion \cite{Brandhuber:2011tv,Bork:2011cj}. Consider the $[n,1\rangle $-supershift of the $\text{N}^k\text{MHV}$ form factor $F_{n,k}$ (for simplicity, we omit the superscript, since we only care about trees in this appendix), where the shift is explicitly written as
\begin{equation}
    |\hat{n}] = |n]+ z |1], \quad 
    |\hat{1}\rangle = |1\rangle - z |n\rangle, \quad
    \hat{\eta}_n = \eta_n + z \eta_1.
\end{equation}
The original form factor $F_{n,k}$ and the shifted form factor $\hat{F}_{n,k}(z)$ are related by the following residue theorem:
\begin{equation}
    F_{n,k}= \text{Res}_{z=0} \frac{\hat{F}_{n,k}(z)}{z} = - \sum_{z_I \neq 0} \text{Res}_{z=z_I} \frac{\hat{F}_{n,k}(z)}{z},
\end{equation}
where $z_I$'s are poles away from $z=0$ and are obtained by solving $\hat{P}_I=\sum_{i\in I}p_i=0$ for a given set of consecutive external momenta including $\hat{1}$ or $\hat{n}$, say $I= \{\hat{1},2,\ldots,j-1\}$. It is well-known that near the pole $z_I$, the residue is evaluated to the gluing of two on-shell parts:
\begin{equation}
\begin{aligned}
    \int d^4 \eta_I \, \hat{A}_{j,k'} \, \frac{1}{P_I} \, \hat{F}_{n-j+2,k-k'-1} = \vcenter{\hbox{\scalebox{.8}{\bcfwAF}}}, \\
    \int d^4 \eta_I \, \hat{F}_{j,k'} \, \frac{1}{P_I} \, \hat{A}_{n-j+2,k-k'-1} =\vcenter{\hbox{\scalebox{.8}{\bcfwFA}}},
\end{aligned}
\end{equation}
where $A_{n,k}$ is the (tree-level) $\text{N}^k\text{MHV}$ superamplitude. Summing over all the residues gives the following recursion formula (in both $n$ and $k$):
\begin{equation}
\begin{aligned}
    F_{n,k}
    & =\left(\sum_{I,k'} \int d^4 \eta_I \, A_{j+1,k'} \, \frac{1}{P_I} \, F_{n-j+1,k-k'-1}\right) + \left( \sum_{I,k'} \int d^4 \eta_I \, F_{j+1,k'} \, \frac{1}{P_I} \, A_{n-j+1,k-k'-1} \right)\\
    & = \sum_{I,k'} \vcenter{\hbox{\scalebox{.8}{\bcfwAF}}} + \sum_{I,k'} \vcenter{\hbox{\scalebox{.8}{\bcfwFA}}}.
\end{aligned}
\end{equation}
Starting from the three-point MHV and $\overline{\text{MHV}}$ amplitudes as well as the two-point form factor, we can obtain all tree-level form factors in principle. The building blocks are as follows:
\begin{equation}
\begin{aligned}
    \vcenter{\hbox{\scalebox{.7}{\diagmhv}}} &  \quad {\Leftrightarrow}\quad 
A_{3,1}(1,2,3) =  \frac{
\delta^{(8)}(\sum_{i}\lambda_{i}\eta_{i})}
{\vev{12}\vev{23} \vev{31}}\,,\\
    \vcenter{\hbox{\scalebox{.7}{\diagantimhv}}} & \quad {\Leftrightarrow}\quad 
A_{3,-1}(1,2,3) = \frac{
\delta^{(4)}([12]\eta_3 + [23] \eta_1 + [31] \eta_2)}
{\sqb{12}\sqb{23}\sqb{31}}\,,\\
    \vcenter{\hbox{\scalebox{.7}{\diagff}}}  &  \quad {\Leftrightarrow}\quad
F_{2,0} (q,\gamma; 1,2) = \frac{
\delta^{(8)}({\gamma + \sum_{i}\lambda_{i}\eta_{i}})}
{\ab{12}\ab{21}}\,.
\end{aligned}
\end{equation}
The simplest case is the MHV form factor. where there is only one contributing diagram.
\begin{equation}
\begin{aligned}
    F_{n,0} & = \vcenter{\hbox{\scalebox{.8}{\bcfwMHV}}}  \\
    & = \int d^4 \eta_I \, A_{3,-1}(\hat{1},2,\hat{P}_I) \, \frac{1}{P_I^2} \, F_{n-1,0}(-\hat{P}_I,3,\ldots,\hat{n}) = \frac{\delta^{(8)}({\gamma + \sum_{i}\lambda_{i}\eta_{i}})}{\ab{12}\ab{23}\cdots\ab{n1}}.
\end{aligned}
\end{equation}
The MHV form factor is similar to the Park-Taylor formula. For NMHV form factor, the recursion formula reads
\begin{equation}
\begin{aligned}
    F_{n,1} = &\vcenter{\hbox{\scalebox{.8}{\bcfwNMHVa}}} + \sum_{j=4}^{n} \vcenter{\hbox{\scalebox{.8}{\bcfwNMHVb}}} \\
     + & \sum_{j=3}^{n-1} \vcenter{\hbox{\scalebox{.8}{\bcfwNMHVc}}} +\vcenter{\hbox{\scalebox{.8}{\bcfwNMHVe}}},
\end{aligned}
\end{equation}
where we write one boundary term explicitly, because it should be treated separately. Performing the Grassmann integration, we get
\begin{equation}
    \vcenter{\hbox{\scalebox{.8}{\bcfwNMHVb}}} = F_{n,0} \, R''_{n2j}, \quad \vcenter{\hbox{\scalebox{.8}{\bcfwNMHVc}}} = F_{n,0} \, R'_{n2j},
\end{equation}
where the $R$-invariants are conventionally drawn as the following on-shell diagrams:
\begin{equation}
\begin{aligned}
    R''_{n2j} = \vcenter{\hbox{\scalebox{.6}{\bcfwNMHVOSb}}}, \quad R'_{n2j} = \vcenter{\hbox{\scalebox{.6}{\bcfwNMHVOSc}}}.
\end{aligned}
\end{equation}
The $R$-invariants are defined as
\begin{equation}
    R^{\square}_{rst}= \hspace{-1em} \vcenter{\hbox{\scalebox{.6}{\Rinv}}} \hspace{-1em} = \frac{\ab{s-1\,s} \ab{t-1 \, t} \, \delta^{(4)}\left( \ab{r|x_{ca}x_{ab}|\theta_{bc}} + \ab{r|x_{cb}x_{ba}|\theta_{ac}} \right)}{x_{ab}^2 \ab{r|x_{cb}x_{ba}|s-1} \ab{r|x_{cb}x_{ba}|s} \ab{r|x_{ca}x_{ab}|t-1} \ab{r|x_{ca}x_{ab}|t}},
\end{equation}
where $\theta_i$ is the dual coordinate of super momentum, \textit{e.g.} $|\theta_{ij}\rangle = |i\rangle \eta_i + \ldots +|j-1\rangle \eta_{j-1}$ (in a periodic sense as in Sec.~\ref{sec:dualcoordinates}), and $\square$ means the definition is for both types of $R$-invariants and the position of the operator leg is encoded in the assignment of dual coordinates, see Fig.~\ref{fig:dualEXT}. In addition, we need to be careful about the boundary term, which is different from the above formula:
\begin{equation}
    R'_{rss} = \hspace{-1em} \vcenter{\hbox{\scalebox{.6}{\RinvRSS}}} \hspace{-1em} = -\frac{\ab{s-1 \, s} \, \delta^{(4)}\left( \ab{r|x_{ca}x_{ab}|\theta_{bc}} + \ab{r|x_{cb}x_{ba}|\theta_{ac}} \right)}{x_{ab}^4 \ab{r|x_{cb}x_{ba}|s-1} \ab{r|x_{ca}x_{ab}|s} \ab{r|x_{ca}x_{bc}|r}}.
\end{equation}
Having the $R$-invariants at hand, the recursion is solved by the following formula:
\begin{equation}
    F_{n,1} = F_{n,0} \left( \sum_{j=2}^{n-2} \sum_{t=j+2}^{n} R''_{njt} + \sum_{j=2}^{n-1} \sum_{t=j}^{n-1} R'_{njt} \right).
\end{equation}
For $\text{N}^2\text{MHV}$, the recursion formula reads
\begin{equation}
\begin{aligned}
    F_{n,2} & = \vcenter{\hbox{\scalebox{.8}{\bcfwNNMHVa}}} + \sum_{j=4}^{n-1}\vcenter{\hbox{\scalebox{.8}{\bcfwNNMHVb}}} + \sum_{j=5}^{n} \vcenter{\hbox{\scalebox{.8}{\bcfwNNMHVc}}}\\
    & + \sum_{j=3}^{n-3} \vcenter{\hbox{\scalebox{.8}{\bcfwNNMHVd}}} + \sum_{j=3}^{n-1} \vcenter{\hbox{\scalebox{.8}{\bcfwNNMHVe}}} + \vcenter{\hbox{\scalebox{.8}{\bcfwNNMHVf}}}.
\end{aligned}
\end{equation}
The simplest case is $n=4$, where we only need to calculate one diagram:
\begin{equation}\label{eq:F42}
\begin{aligned}
    F_{4,2} & = \vcenter{\hbox{\scalebox{.8}{\bcfwNNMHVfourPT}}} = \int d^4 \eta_I \, F_{3,1}(\hat{1},2,\hat{P}_I) \, \frac{1}{P^2} \, A_{3,0}(-\hat{P}_I,3,\hat{4})\\
    & = \underbrace{\int d^4 \eta_I \, F_{3,0}(\hat{1},2,\hat{P}_I) \, \frac{1}{P^2} \, A_{3,0}(-\hat{P}_I,3,\hat{4})}_{F_{4,0} \times R'_{423}} \times \vcenter{\hbox{\scalebox{.6}{\RinvNNMHVfourPTa}}}.
\end{aligned}
\end{equation}
In the $R$-invariant on the right, we need to replace the corresponding spinor by $|\hat{1}\rangle$ and $|\hat{P}_I\rangle$. Note that they appear homogeneously in the numerator and denominator of the $R$-invariants, so we are free to multiply/divide them by some factor. Consider $|\hat{P}_I\rangle [\hat{P}_I 1]\ab{14} = (3+4)1|4\rangle$, thus we can make the replacement: $|\hat{P}_I\rangle \rightarrow (3+4)1|4\rangle$. For $|\hat{1}\rangle$, we need to solve $z_I$ first:
\begin{equation}\label{eq:rep1}
    0=\hat{P}_I^2 =(p_3 + p_{\hat{4}})^2 = 2\, p_3 \cdot p_{\hat{4}} = \langle 4 |3| \hat{4} ] = \langle 4 |3| 4 ] + z_I \langle 4 |3| 1] \Rightarrow z_I = - \frac{\langle 4 |3| 4 ]}{\langle 4 |3| 1]},
\end{equation}
and then we get
\begin{equation}\label{eq:rep2}
    | \hat{1} \rangle = | 1 \rangle - z_I | 4 \rangle = \frac{| 1 \rangle [1|3|4\rangle + | 4 \rangle [4|3|4\rangle}{[1|3|4\rangle} = \frac{(1+4)3|4\rangle}{[1|3|4\rangle}.
\end{equation}
And we can replace $| \hat{1} \rangle$ by $(1+4)3|4\rangle$. Above all, we now have:
\begin{equation}\label{eq:F42final}
    F_{4,2} = F_{4,0} \times \vcenter{\hbox{\scalebox{.6}{\RinvNNMHVfourPTb}}} \times \left. \vcenter{\hbox{\scalebox{.6}{\RinvNNMHVfourPTa}}} \right|_{\substack{| \hat{1} \rangle \rightarrow (1+4)3|4\rangle\\|\hat{P}_I\rangle \rightarrow (3+4)1|4\rangle}},
\end{equation}
where we have a modified $R$-invariant for the first time. Now we have seen all the ingredients to calculate all tree-level form factors by the BCFW recursion. For higher point $n$ and Grassmann degree $k$, the calculation is similar to Eq.~\eqref{eq:F42}: We first factor out the MHV part of the on-shell form factor/amplitude, which together integrate to $F_{n,0}$ multiplied by an $R$-invariant. Then we need to solve $z_I$ and determine the corresponding replacement of the $R$-invariants as in Eq.~\eqref{eq:rep1} and \eqref{eq:rep2}. The final result will always has the form of $F_{n,0}$ times a summation of product of several (modified) $R$-invariants.

Having the BCFW results at hand, the squaring procedure is straightforward. One way is to expand the result in terms of the Grassmann variables $\eta$ to get all possible helicity configurations and sum over the square of all components. This calculation can be cumbersome since the number of the components is quite big. The other way is to directly manipulate the $R$-invariants. We first define the ``ratio'' of the form factor $M_{n,k} = F_{n,k} / F_{0,k}$. Then $M_{n,k}$ contains only $R$-invariants. The ``ratio'' of the form factor square (see Eq.~\eqref{eq:4ptTreeMandel} and \eqref{eq:4ptTreeDual}) is then
\begin{equation}
    \frac{\FFint_{n}}{\FFint_{n,0}} = \left(\sum_{k=0}^{n-2} M_{n,k} M_{n,n-2-k} \right) / M_{n,n-2}.
\end{equation}
The numerator on the r.h.s. is a sum of product of $R$-invariants, which can be conveniently calculated by the \texttt{Mathematica} package \cite{Bourjaily:2023uln}. This numerator is also proportional to $M_{n,n-2}$ (\textit{i.e.} the ratio of $\overline{\text{MHV}}$), thus all the Grassmann variables cancel on the r.h.s. as expected. Thus, in practice, we can extract an arbitrary component of the numerator (this can be done by the function \texttt{component} in previously mentioned package) and divide by the same component in $M_{n,n-2}$. Then we will have an expression in terms of spinor helicity. We have numerically checked that our results obtained in the main text are consistent with those obtained by BCFW recursion described in this appendix.

\bibliographystyle{JHEP}
\bibliography{inspire.bib}

\end{document}